\documentclass[notitlepage]{revtex4-1}
\usepackage{amsmath}
\usepackage{bm}
\usepackage{natbib}
\usepackage{graphicx}
\usepackage{hyperref}
\usepackage{color,soul}

\begin{document}
%front matter
\title{High-order simulations of isothermal flows using the local anisotropic basis function method (LABFM)}

%% UNCOMMENT IF REVTEX
\author{J. R. C. King}\email{jack.king@manchester.ac.uk}\affiliation{Department of Mechanical, Aerospace and Civil Engineering, The University of Manchester, Manchester, UK}
\author{S. J. Lind}\affiliation{Department of Mechanical, Aerospace and Civil Engineering, The University of Manchester, Manchester, UK}

%% UNCOMMENT IF ELSARTICLE
%\author[1]{J. R. C. King\corref{cor1}}\ead{jack.king@manchester.ac.uk}
%\cortext[cor1]{Corresponding author}\address[1]{Department of Mechanical, Aerospace and Civil Engineering, The University of Manchester, Manchester, UK}

\begin{abstract}

Mesh-free methods have significant potential for simulations of flows in complex geometries, with the difficulties of domain discretisation greatly reduced. However, many mesh-free methods are limited to low order accuracy. In order to compete with conventional mesh-based methods, high order accuracy is essential. The Local Anisotropic Basis Function Method (LABFM) is a mesh-free method introduced in \emph{King et al., J. Comput. Phys. 415:109549 (2020)}, which enables the construction of highly accurate difference operators on disordered node discretisations. Here, we introduce a number of developments to LABFM, in the areas of basis function construction, stencil optimisation, stabilisation, variable resolution, and high order boundary conditions. With these developments, direct numerical simulations of the Navier Stokes equations are possible at extremely high order (up to $10^{th}$ order in characteristic node spacing internally). We numerically solve the isothermal compressible Navier Stokes equations for a range of geometries: periodic and channel flows, flows past a cylinder, and porous media. Excellent agreement is seen with analytical solutions, published numerical results (using a spectral element method), and experiments. The potential of the method for direct numerical simulations in complex geometries is demonstrated with simulations of subsonic and transonic flows through an inhomogeneous porous media at pore Reynolds numbers up to $Re_{p}=968$.

\end{abstract}

\maketitle
%end front matter

\section{Introduction\label{intro}}

%% DNS
In the context of computational fluid dynamics, a desirable property of a numerical scheme is that it converges quickly with resolution refinement towards the physical system it is designed to represent. Denoting a length scale of the discretisation as $s$, a ``high order'' scheme is generally considered to be one where the dominant errors in the discrete system scale with $s^{m}$ and $m>2$. A consistent high order scheme provides reassurance that simulation results may be relied upon, which is particularly important in an engineering design context. Furthermore, high order methods provide in equivalent accuracy at coarser resolutions than their low order counterparts, hence reducing computational costs. For simple geometries, pseudo-spectral and high order finite difference (FD) methods provide a high degree of accuracy, and are extremely efficient, and as such have dominated the field of direct numerical simulations (DNS) since its inception. However, such methods are limited to simple geometries, and the development of consistent, high order numerical methods capable of simulating flows in complex geometries is a major goal for numerical modellers.

%IBM
Immersed Boundary Methods (IBM) in conjunction with FD schemes can be used to model complex geometries on a simple structured mesh~\cite{mittal_2005}, and have been widely used in the field of fluid-structure interaction~\cite{griffith_2020}. However, such an approach is not suited to the incorporation of spatially varying resolution (e.g. near solid boundaries), and the use of IBM for DNS is limited by difficulties in the accurate satisfaction of boundary conditions, which in most IBM codes are limited to first or second order~\cite{stein_2017}. Notwithstanding this, the approach has been used in conjunction with low order schemes for complex geometries, including for example cylinder-array representations of porous media~\cite{liu_2020}, and more realistic porous geometries~\cite{finn_2013,jin_2017}. Recently, IBM have been used in conjunction with high order FD schemes for simulations of turbulent combustion~\cite{rauch_2018}, although the imposition of boundaries relies on second order interpolation, and the approach at high order has so far been limited to relatively simple geometries. Finite volume (FV) methods provide an alternative approach, and used with a structured mesh and high order reconstruction schemes, they can yield excellent accuracy and conservation properties, although again with limited geometric flexibility. Geometric flexibility can be obtained by constructing FV schemes on an unstructured (body fitted) mesh, though this flexibility is typically traded for a loss in accuracy, with schemes generally second order. Furthermore, skewed mesh elements (which are inevitable in a complex geometry), can have a substantial detrimental effect on numerical accuracy~\cite{finn_2013}. 

The Spectral Element Method (SEM)~\cite{sem_1989} provides high order accuracy and geometric flexibility, by using an unstructured mesh, with local high order polynomial or spectral interpolation within each mesh element. SEM has been widely applied to problems in fluid dynamics, seismic wave propagation, and ocean modelling. A comprehensive overview of spectral element methods is given in~\cite{sem_2013}. As with FV methods, accuracy deteriorates in the presence of skewed mesh elements, and the generation of a high quality mesh is challenging and costly in complex geometries. A number of related methods have been developed for complex geometries, which solve the governing equations in their weak form, including high-order discontinuous Galerkin methods~\cite{shahbazi_2007}, cut-FEM~\cite{burman_2015}, and conformal decomposition FEM~\cite{fries_2018}. In situations where the resolution of both hydrodynamic and acoustic flow features is required, these weak form methods can be prohibitively expensive, as their implicit or semi-implicit nature necessitates the solution of a large sparse linear system every time-step.

%% application
Our interest here is in moderate to high Reynolds number flows through complex geometries, and in particular porous media, an area where accurate simulations are key, as experimental studies of pore-scale flows are often impossible (without contaminating the flow), and the applicability of statistical turbulence models is questionable~\cite{jin_2017}. Such flows are found in a variety of applications, including filtration systems, porous burners, hydrogen and syngas production, and pebble bed reactors for nuclear power generation~\cite{wood_2020}. However, pore-resolved simulations present a challenge for the mesh-based methods described above, where the generation of a high quality body-fitted mesh is an extremely time-consuming process (often taking longer than the simulation itself~\cite{wood_2020}), and pore-scale simulations are often limited to idealised representations of porous media (e.g.\cite{jin_2015,kuwata_2017,chu_2019,liu_2020}). 

%Mesh-free methods...
Mesh-free (or meshless) numerical methods discretise the computational domain into a disordered set of nodes, particles, or collocation points. No information on the topology of the discretisation is required; derivatives or fluxes are evaluated using operators based solely on the relative positions of the nodes. For a broad overview of mesh-free methods, we refer the reader to~\cite{li_review_2002,sahil_2018}. With no information on nodal connectivity needed, mesh-free methods largely obviate the difficulties of domain discretisation. Algorithms to fill surfaces and volumes with a set of nodes (e.g.~\cite{fornberg_2015a}) are computationally cheap and easily automated, and mesh-free methods are well suited to problems involving complex geometries. In an Eulerian formulation, high-order mesh-free methods may provide the geometric flexibility of SEM and other body-conforming mesh-based methods, but with explicit time-stepping, thus reducing the costs of compressible flow simulations, and leading to simpler parallelisation. High order mesh-free methods are therefore attractive, and even in Eulerian formulations have the potential to transform the field of DNS, enabling detailed studies of fundamental flow phenomena in highly complex geometries.

%LABFM
With this motivation, in~\cite{king_2020} we introduced the Local Anisotropic Basis Function method (LABFM) - a novel mesh-free method for generating high order spatial derivative operators. Derivative operators are constructed from the weighted sum (over the computational stencil) of field value differences, where the weights are obtained from a linear combination of anisotropic basis functions (ABFs). The combination of ABFs is constructed by solving a local linear system for each stencil to ensure polynomial consistency of the desired order. The method was applied to a range of prototypical PDEs, and a simple unbounded incompressible flow problem, showing convergence up to $8^{th}$ order. We provided a detailed overview of the state of the art in mesh-free methods in~\cite{king_2020}, to which we refer interested readers. Compared with other mesh-free methods, LABFM provides much greater accuracy than Smoothed Particle Hydrodynamics (SPH), which in typical use (i.e. uncorrected and Lagrangian) is not even zero order consistent. Furthermore, due to the construction of ABFs, LABFM yields higher order at reduced computational cost compared with the various consistency corrections to SPH (e.g.~\cite{bonet_lok,zhang_2004,asprone_2010,asprone_2011,sibilla_2015}), the Reproducing Kernel Particle Method (RKPM)~\cite{liu_1995,liu_1995b}, and the generalised finite difference method (GFDM)~\cite{benito_2007,gavete_2017,suchde_2018}, which although capable of high order, are generally utilised in their second order formulations in practice. In terms of accuracy, LABFM is comparable with Radial Basis Function (RBF) finite difference methods~\cite{wright_2003,fornberg_2015}, but with smaller stencils for a given level of consistency, or smaller linear systems, depending on the formulation, again resulting in reduced computational costs. Compact moving least squares (CMLS)~\cite{trask_2016} and generalised moving least squares (GMLS)~\cite{trask_2017} bear similarities to LABFM, although the philosophy of the derivation differs: in GMLS, the polynomial reconstruction of a function a obtained by formulating the interpolation as an optimisation problem, with derivative operators then obtained by taking diffuse derivatives of the reconstruction. In GMLS and CMLS, the local consistency matrices in GMLS are symmetric, with Taylor monomials most commonly used as the basis, whilst the systems in LABFM are generally asymmetric. GMLS has been used to study Stokes flow problems in an implicit divergence-free-preserving formulation~\cite{trask_2018}, including the implementation of dynamic refinement schemes through node/particle splitting techniques~\cite{hu_2019}.

In this paper, we build on~\cite{king_2020} and introduce two new developments that greatly increase the generality of LABFM. These are (i) new stabilisation techniques for dissipation of high wavenumber numerical instabilities (a common issue in all high order collocated schemes) and (ii) high order boundary conditions for complex geometries, including no-slip, inflow, and outflow conditions. Our particular area of interest is in the weakly compressible flow regime with Mach numbers typically in the range $0$ to $0.2$, and accordingly the new boundary conditions take special care in the treatment of acoustic energy and vorticity in the domain. Flow variables at boundaries are treated in a physically consistent way with negligible unphysical behaviour (e.g. reflections at outflow) with boundary discretisations keeping high ($4^{th}$ order) spatial accuracy. 

%% Boundary conditions.
For the compressible Navier-Stokes equations, boundary conditions are often implemented using characteristic based formulations~\cite{poinsot_1992}. Here the governing equations are decomposed to obtain evolution equations for the characteristic waves (describing acoustic, entropy and vorticity propagation). Numerical boundary conditions are applied by specifying the amplitude of incoming waves (at the boundary) such that physical boundary conditions are satisfied in a way that is consistent with the Navier-Stokes equations. This approach is widely used in mesh-based methods (e.g.~\cite{yoo_2005,yoo_2007,fosso_2012,motheau_2017}), but is less prevalent in mesh-free methods. One reason for this is that characteristic boundary conditions require one-sided derivative operators at the boundary, which are not possible (or of very limited accuracy) for many mesh-free methods, where low-order extrapolation approaches are preferred (e.g.~\cite{adami_2012,tafuni_2018}). A non-reflecting boundary condition based on characteristc waves has been developed in~\cite{wang_2019} for a Lagrangian SPH scheme. Here, in the absence of one-sided derivative operators, a Lagrangian interpolation is used to extrapolate properties to a set of dummy particles, and the accuracy is limited to the accuracy of the underlying SPH method. Furthermore, the approach is only applicable to non-reflecting outflow boundaries. High order boundary conditions have been developed for mesh-free methods for incompressible flows. Recent work in~\cite{nasar_2020} introduced high order extrapolations for the imposition of Neumann and Dirichlet boundary conditions in an Eulerian SPH scheme. Although spatially high order, the approach is used in a fractional step algorithm, for which the temporal accuracy with which the continuity constraint is satisfied is limited to first order~\cite{rempfer_2006}.
RBF-FD is one mesh-free method where one-sided derivative operators are possible~\cite{bayona_2017}, although this fact doesn't appear to have been exploited for the application of characteristic based boundary conditions, with most RBF-FD studies involving fluid simulations focussed on the incompressible Navier-Stokes equations (e.g.~\cite{javed_2014,abba_2020}). The developments herein represent the first implementation of characteristic based boundary condition formulations within a high-order meshless framework. 

%% Collocated methods and filtering.
With regard to stabilisation, in the absence of specific treatments, high order collocated methods (including LABFM) for solving the Navier Stokes equations admit spatial instabilities close to the Nyquist wavenumber of the discretisation. In pseudo-spectral and finite difference methods, these are typically dealt with by dealiasing~\cite{orszag_1971} or filtering~\cite{kennedy_1994,brandenburg_2003}. For low order methods, the discrete Laplacian operator often contains sufficient numerical diffusivity to suppress the development of these instabilities. For Lagrangian methods, such as SPH, these instabilities do not arise, as the quadratically non-linear advection terms are not explicitly calculated, and upwinding is effectively included through the Lagrangian motion of the collocation points. With many mesh-free methods either Lagrangian or of low order, the issue has not been particularly prominent to date. Recently, high order has been demonstrated in an Eulerian SPH framework using modified kernel functions~\cite{lind_2016,nasar_2020}, with no stabilisation techniques required. Although high order convergence was observed for a range of resolutions in~\cite{lind_2016,nasar_2020}, the schemes are formally zero order consistent on general node distributions. Instabilities were noted in~\cite{nasar_2020}, but their suppression required no special treatment beyond careful adjustment of the time step. In this case we postulate that the conditional stability benefits from increased numerical dissipation through low-order time stepping and the lack of formal consistency of the Laplacian operators, which are based on the Morris operator widely used in SPH~\cite{morris_1997}. The problem of high wavenumber instabilities in mesh-free methods has been circumvented by constructing mesh-free methods in a finite volume formulation, using local RBF methods~\cite{shu_2005,dehghan_2018}, and an SPH derived approach~\cite{hopkins_2015}. However, issues around the definition of cell volumes in a mesh-free context remain an obstacle to convergence rates above second order. In the landscape of meshfree methods, RBF-FD and LABFM are outliers, and with stability and accuracy properties more akin to high order finite differences than other mesh-free methods, some form of stabilisation is necessary for advection-dominated problems. For RBF-FD methods, the approach to stabilisation has generally been to move the filtering operation inside the governing equations~\cite{fornberg_2011}. By adding a small amount of hyperviscosity to the governing equations, the generation of high wavenumber modes can be suppressed, with minimal effect on the (low wavenumber) physical components of the flow. This approach has been widely used in the RBF-FD community, with recent developments focussed on automatic parameter setting for advection-dominated problems on surfaces and manifolds~\cite{shankar_2018,shankar_2020}. In~\cite{king_2020} we followed this approach with LABFM to enable solutions of Burgers' equation at higher Reynolds numbers. In this work stabilisation is achieved through filtering variables at each time step and has yielded a robust procedure capable of stabilising simulations across a range of flows without the need for parameter tuning.

In addition to the above, we also present several smaller improvements to the methodology in~\cite{king_2020}, including the use of orthogonal basis polynomials for improved matrix conditioning, the inclusion of spatially varying resolution, and a simple but effective technique for stencil optimisation based on errors in the local consistency matrices. All of the above developments are tested over the isothermal Navier Stokes equations solved in an Eulerian frame for a wide range of internal flow test cases (steady, unsteady and implusive, and for a variety of boundary conditions) for Reynolds numbers up to $1000$, and Mach numbers ranging from $0.2$ to the incompressible limit. We conclude with a geometrically challenging test problem of flow through an inhomogeneous porous media, which although not quantitatively verifiable, demonstrates the potential of the method for high order simulations in complex geometries.

The remainder of the paper is arranged as follows. In Section~\ref{labfm} we provide a brief recap of LABFM. In Section~\ref{abf} we introduce a new method for constructing ABFs with significant advantages over our original approach. Section~\ref{stencil} details a stencil optimisation procedure to improve the accuracy and reduce the costs of the method. In Section~\ref{varres} the ability of LABFM to handle spatially varying resolution is demonstrated. In Section~\ref{filters} we introduce a filtering technique which can be used to stabilise numerical simulations the Navier Stokes equations. Section~\ref{ns} details the governing equations for isothermal compressible fluids, and describes our boundary condition formulation. Section~\ref{numres} contains numerical results providing validation of the method for simple flows, and a demonstration of the potential for simulating flows in complex geometries with LABFM. Section~\ref{conc} is a summary of conclusions.

\section{The Local Anisotropic Basis Function Method}\label{labfm}

\begin{figure}
\includegraphics[width=0.5\textwidth]{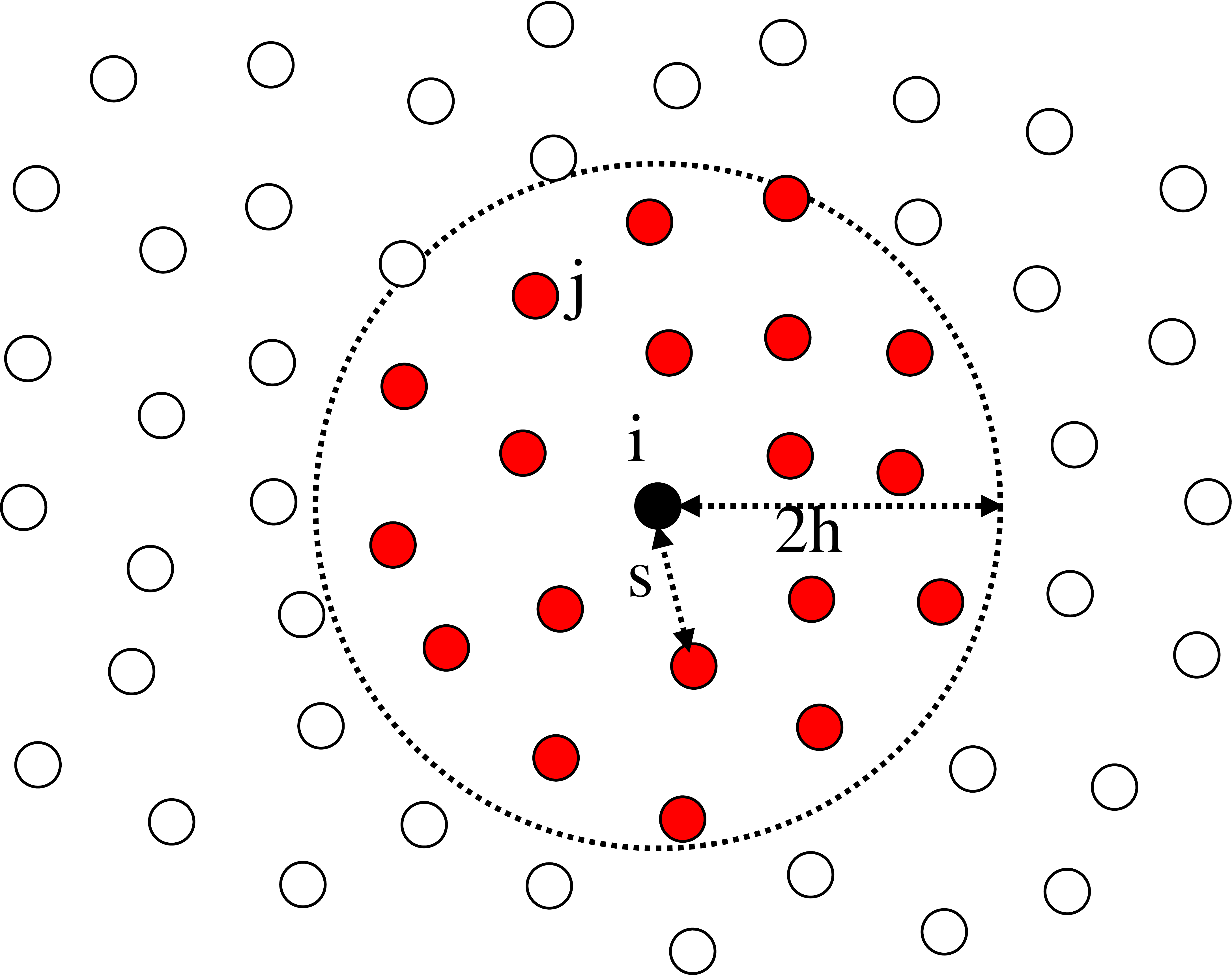}
\caption{An illustration of the computational stencil.\label{fig:stencil}}
\end{figure}

For a detailed derivation and analysis of the Local Anisotropic Basis Function method we refer the reader to~\cite{king_2020}. Here we present only the details required for reproduction the method. The domain is discretised with a quasi-uniform point cloud of $N$ nodes. In this work we consider only two-dimensional problems; although an extension to three dimensions is straightforward, three-dimensional simulations are reserved for a future study. Each node $i\in\left[1,N\right]$ has position $\bm{r}_{i}=\left(x_{i},y_{i}\right)^{T}$, a distribution lengthscale $s_{i}$, and characteristic stencil lengthscale $h_{i}$. $s_{i}$ is the average node spacing around node $i$, analagous to the grid spacing in a mesh-based finite difference scheme. The difference between properties at two nodes is denoted $\left(\cdot\right)_{ji}=\left(\cdot\right)_{j}-\left(\cdot\right)_{i}$. Each node has a computational stencil containing all $\mathcal{N}_{i}$ nodes $j$ such that $\lvert\bm{r}_{ji}\rvert\le2h_{i}$. Figure~\ref{fig:stencil} illustrates the node distribution and stencil. The number of neighbours $\mathcal{N}_{i}$ is related to the distribution lengthscale $s_{i}$ by $\mathcal{N}_{i}\approx4\pi\left(h_{i}/s_{i}\right)^{2}$. We use the ratio $h_{i}/s_{i}$ to characterise the stencil size. To improve the clarity of exposition and accomodate non-uniform resolutions we have made some minor changes to the notation compared with~\cite{king_2020}: $\delta{r}$ in~\cite{king_2020} corresponds to $s_{i}$ here, and where $k$ represented the order of LABFM in~\cite{king_2020}, here it is denoted by $m$. We use the distribution lengthscale $s_{i}$ (made dimensionless with a domain lengthscale) as a measure of the resolution.  

The following procedure is applied to every node $i$, resulting in a set of weights which define the difference operators. For the Eulerian schemes used herein, this is done as a pre-processing step. For Lagrangian schemes, this would be done every time-step. We introduce a vector of monomials
\begin{equation}\bm{X}_{ji}=\begin{bmatrix}x_{ji},&y_{ji},&\frac{x_{ji}^{2}}{2},&x_{ji}y_{ji},&\frac{y_{ji}^{2}}{2},&\frac{x_{ji}^{3}}{6},&\frac{x_{ji}^{2}y_{ji}}{2},&\frac{x_{ji}y_{ji}^{2}}{2},&\frac{y_{ji}^{3}}{6},&\frac{x_{ji}^{4}}{24},\dots\frac{y_{ji}^{m}}{m!}\end{bmatrix}^{T},\end{equation}
and a vector operator of partial derivatives as
\begin{equation}\bm{D}\left(\cdot\right)=\begin{bmatrix}\frac{\partial\left(\cdot\right)}{\partial{x}},&\frac{\partial\left(\cdot\right)}{\partial{y}},&\frac{\partial^{2}\left(\cdot\right)}{\partial{x}^{2}},&\frac{\partial^{2}\left(\cdot\right)}{\partial{x}\partial{y}},&\frac{\partial^{2}\left(\cdot\right)}{\partial{y}^{2}},&\frac{\partial^{3}\left(\cdot\right)}{\partial{x}^{3}},&\frac{\partial^{3}\left(\cdot\right)}{\partial{x}^{2}\partial{y}},&\frac{\partial^{3}\left(\cdot\right)}{\partial{x}\partial{y}^{2}},&\frac{\partial^{3}\left(\cdot\right)}{\partial{y}^{3}},&\frac{\partial^{4}\left(\cdot\right)}{\partial{x}^{4}},\dots,\frac{\partial^{m}\left(\cdot\right)}{\partial{y}^{m}}\end{bmatrix}^{T},\end{equation}
where $m$ is the desired order of polynomial reproduction. Note the vector $\bm{X}_{ji}$ has length $n=\left(m^{2}+3m\right)/2$. We define a general discrete operator, operating on node $i$:
\begin{equation}L^{d}_{i}\left(\cdot\right)=\displaystyle\sum_{j}\left(\cdot\right)_{ji}w^{d}_{ji}\label{eq:general_do}\end{equation}
where $d$ identifies the specific partial derivative(s) being approximated, and $w^{d}_{ji}$ are the set of weights. Here (and throughout), the sum over $j$ is over all $\mathcal{N}_{i}$ nodes which are neighbours of $i$. Equation~\eqref{eq:general_do} approximates $\left.\bm{C^{d}}\cdot\bm{D}\left(\cdot\right)\right\rvert_{i}$ where $\bm{C^{d}}$ is a vector pointing to the appropriate derivatives for a given $d$. For example, we are often interested in estimating the gradients and Laplacian of a function, for which cases we set $d=x$, $d=y$, and $d=L$, giving
\begin{equation}\bm{C^{d}}=\begin{cases}\begin{bmatrix}1,&0,&0,&0,&0,&0,&0,&0\dots\end{bmatrix}^{T}&\text{if }d=x\\
\begin{bmatrix}0,&1,&0,&0,&0,&0,&0,&0\dots\end{bmatrix}^{T}&\text{if }d=y\\\begin{bmatrix}0,&0,&1,&0,&1,&0,&0,&0\dots\end{bmatrix}^{T}&\text{if }d=L.\end{cases}\end{equation}
We set $w^{d}$ equal to the weighted sum of a series of Anistropic Basis Functions (ABFs) $W_{ji}=W\left(\bm{r}_{ji}/h_{i}\right)$, writing
\begin{equation}w^{d}_{ji}=\bm{W}_{ji}\cdot\bm{\Psi^{d}}_{i}=W^{1}_{ji}\Psi^{d}_{i,1}+W^{2}_{ji}\Psi^{d}_{i,2}+W^{3}_{ji}\Psi^{d}_{i,3}+W^{4}_{ji}\Psi^{d}_{i,4}+\dots+W^{n}_{ji}\Psi^{d}_{i,n}\label{eq:w_abf}\end{equation}
where the vector of basis functions is $\bm{W}_{ji}=\left[W^{1}_{ji},W^{2}_{ji},W^{3}_{ji},\dots,W^{n}_{ji}\right]^{T}$ and the vector of weights is $\bm{\Psi^{d}}_{i}=\left[\Psi^{d}_{i,1},\Psi^{d}_{i,2},\Psi^{d}_{i,3},\dots,\Psi^{d}_{i,n}\right]^{T}$. We construct the matrix $\bm{M}_{i}$ as
\begin{equation}\bm{M}_{i}=\displaystyle\sum_{j}\bm{X}_{ji}\otimes\bm{W}_{ji},\label{eq:disc_form}\end{equation}
and then obtain the weights $\bm{\Psi^{d}}_{i}$ by solving the system
\begin{equation}\bm{M}_{i}\bm{\Psi^{d}}_{i}=\bm{C^{d}}.\label{eq:lsys}\end{equation}
Having solved~\eqref{eq:lsys} (in our implementation we do so through LU decomposition using the LAPACK package~\cite{lapack_1999}) we use~\eqref{eq:w_abf} to obtain the weights $w^{d}_{ji}$. The resulting operator $L$ yields polynomial reproduction of order $m$: the leading order error is $\mathcal{O}\left(h_{i}^{m+1-l}\right)$, where $l$ is the order of the derivative being approximated. Provided the resolution $s$ is refined such that $h/s$ remains constant, this translates to errors of order $\mathcal{O}\left(s_{i}^{m+1-l}\right)$. In~\cite{king_2020}, we defined ABFs as the partial derivatives of a fundamental RBF, such that
\begin{equation}\bm{W}_{ji}=\left.\bm{D}\psi\left(\lvert\bm{r}_{ji}\rvert/h_{i}\right)\right\rvert_{j}\label{eq:originalabfs},\end{equation}
where the partial derivatives are evaluated with respect $\bm{r}_{j}$. Although this approach provided very good results in~\cite{king_2020}, in Section~\ref{abf} we present details of an alternative method for constructing the ABFs, which has distinct advantages. We also note here a modification which improves the condition number of $\bm{M}$, by scaling each row with a power of $h_{i}$. We introduce a vector containing powers of $h$
\begin{equation}\bm{\mathcal{H}}_{i}=\begin{bmatrix}h_{i}^{-1},&h_{i}^{-1},&h_{i}^{-2},&h_{i}^{-2},&h_{i}^{-2},&h_{i}^{-3},&h_{i}^{-3},&h_{i}^{-3},&h_{i}^{-3},\dots,&h_{i}^{-m}\end{bmatrix}^{T},\end{equation}
and define $\widetilde{\bm{X}}_{ji}=\bm{X}_{ji}\cdot\bm{\mathcal{H}}_{i}$ and $\widetilde{\bm{C^{d}}}_{i}=\bm{C^{d}}\cdot\bm{\mathcal{H}}_{i}$. Replacing $\bm{X}_{ji}$ and $\bm{C^{d}}$ with $\widetilde{\bm{X}}_{ji}$ and $\widetilde{\bm{C^{d}}}_{i}$ in~\eqref{eq:disc_form} and~\eqref{eq:lsys}, and noting that the exact solution of the linear system is unchanged, results in an improvement of the condition number of $\bm{M}$ by several orders of magnitude.

\section{New ABFs based on orthogonal polynomials}\label{abf}

\begin{figure}
\includegraphics[width=0.49\textwidth]{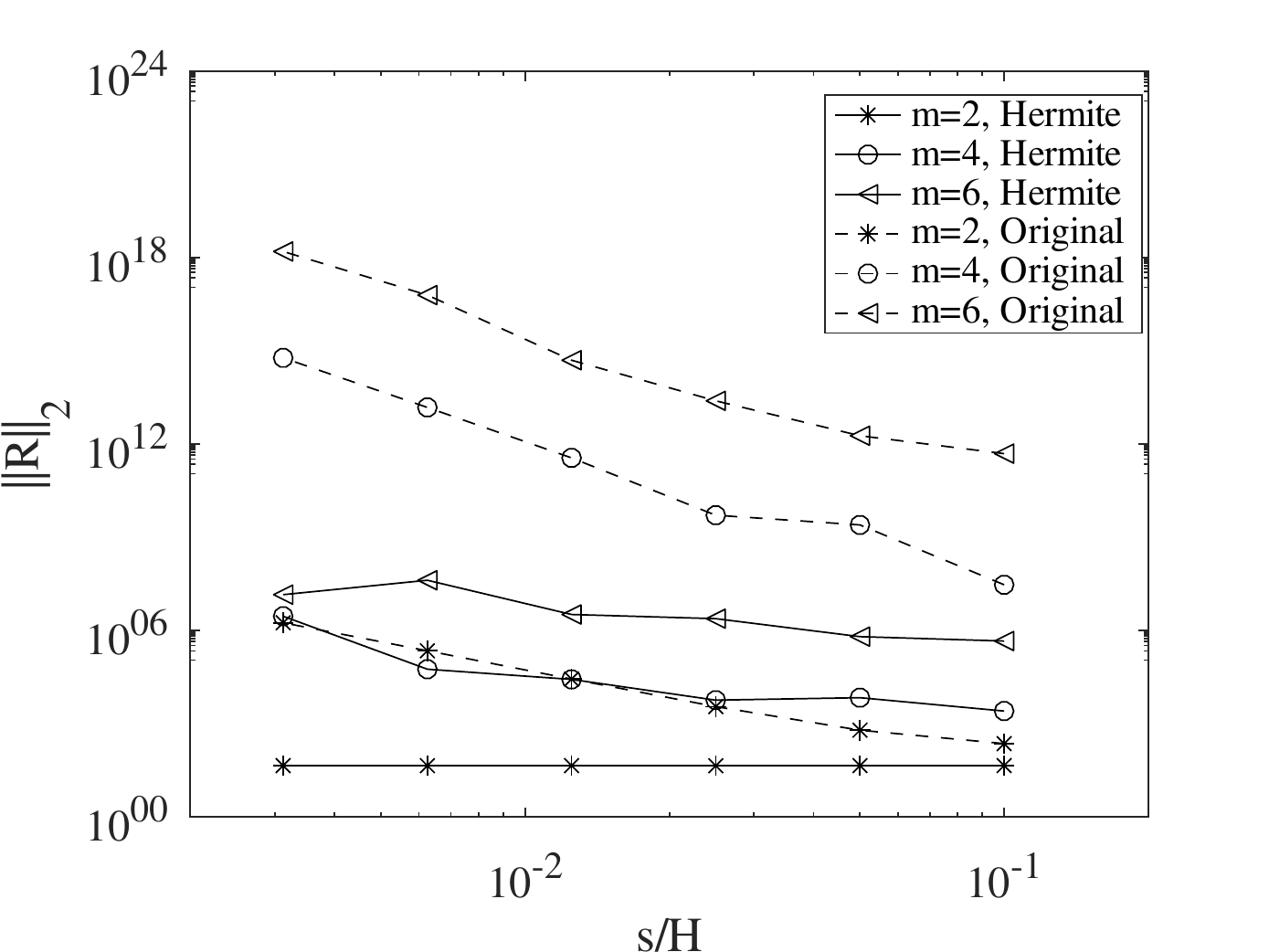}
\caption{Variation of the $L_{2}$ norm of the condition number $R$ of the matrix $\bm{M}$ with resolution $s/H$, for several orders of LABFM (values of $m$), for original ABFs as in~\cite{king_2020} (dashed lines) and Hermite ABFs (solid lines). The Hermite ABFs are multiplied by a Wendland C2 RBF. The $L_{2}$ norm is calculated over a square domain of side length $H$, with a distribution noise level of $\varepsilon/s=0.5$.\label{fig:cond}}
\end{figure}

The ABFs introduced in~\cite{king_2020} and defined by~\eqref{eq:originalabfs} (which we hereafter refer to as ``original'' ABFs) do not form an orthogonal basis. Whilst for the two-dimensional cases explored in~\cite{king_2020} this did not pose any problems, the original ABFs cannot in general guarantee linear independence of the rows of $\bm{M}$. The condition number of $\bm{M}$ is large generally (see dashed lines in Figure~\ref{fig:cond}), with a larger condition number resulting in larger residual errors in $\bm{\Psi^{d}}$, and decreasing the accuracy of the operator $L$. An example of where this results in the method breaking down is when LABFM is extended to three-dimensions, with $m=4$. Here $\bm{M}$ has size $n=34$, and rank $n-1=33$, regardless of the node distrubtion. This can be shown to be the case in the limit of large $\mathcal{N}$, when the sum in~\eqref{eq:disc_form} becomes a continuous integral, either by algebra or numerical experiment (to within machine precision). 

An alternative approach to~\eqref{eq:originalabfs} is to construct ABFs from classical orthogonal polynomials. In this work we use the bivariate Hermite polynomials as the basis for our ABFs, combined with a radial basis function (RBF). Where the $q$-th element of $\bm{X}_{ji}$ is proportional to $x_{ji}^{a}y_{ji}^{b}$, we define the $q$-th ABF as
\begin{equation}W_{ji}^{q}=\frac{\psi\left(\lvert\bm{r}_{ji}\rvert/h_{i}\right)}{\sqrt{2^{a+b}}}H_{a}\left(\frac{x_{ji}}{h_{i}\sqrt{2}}\right)H_{b}\left(\frac{y_{ji}}{h_{i}\sqrt{2}}\right),\label{eq:hermite}\end{equation}
in which $H_{a}$ is the $a$-th order univariate Hermite polynomial (of the physicists kind), generated from
\begin{equation}H_{a}\left(z\right)=\left(-1\right)^{a}e^{z^{2}}\frac{d^{n}}{dz^{n}}e^{-z^{2}},\end{equation}
and $\psi$ is an RBF.
Unless explicitly stated otherwise, we use the Wendland C2 RBF as detailed in~\cite{king_2020,dehnen_aly}, although other RBFs (or even a constant RBF $\psi=1$) are acceptable. We refer to ABFs generated by~\eqref{eq:hermite} as Hermite ABFs.

The Hermite polynomials used in~\eqref{eq:hermite} form an orthogonal basis, and hence the rows of $\bm{M}$ are linearly independent. This results in significant improvements in the condition number of $\bm{M}$ compared with the original ABFs. Figure~\ref{fig:cond} shows the variation of the condition number $R$ of the matrix $\bm{M}$ with resolution for original (dashed lines) and Hermite (solid lines) ABFs, for $m=2,4,6$. The condition number is several orders of magnitude smaller for the Hermite ABFs than the original ABFs. The increase in condition number with resolution refinement is greater for original ABFs. This represents a significant advantage of Hermite ABFs over original ABFs. An additional benefit of Hermite ABFs over original ABFs is the reduced cost of computation, which although less relevant for the Eulerian schemes used in this work, may become significant in a Lagrangian formulation of LABFM. The Hermite ABFs in Figure~\ref{fig:cond} used the Wendland C2 RBF for $\psi$. We note that if we use a lower order RBF (for example, a constant, conic or quadratic RBF), the condition number increases by several orders of magnitude. Although in this work we use Hermite ABFs, we have also tested the method with Legendre and Laguerre polynomials. We find negligible difference in the condition number of $\bm{M}$ and the accuracy of $L^{d}$ if the Legendre or Laguerre polynomials (with appropriately scaled arguments) are used in place of Hermite polynomials in~\eqref{eq:hermite}.

\begin{figure}
\includegraphics[width=0.49\textwidth]{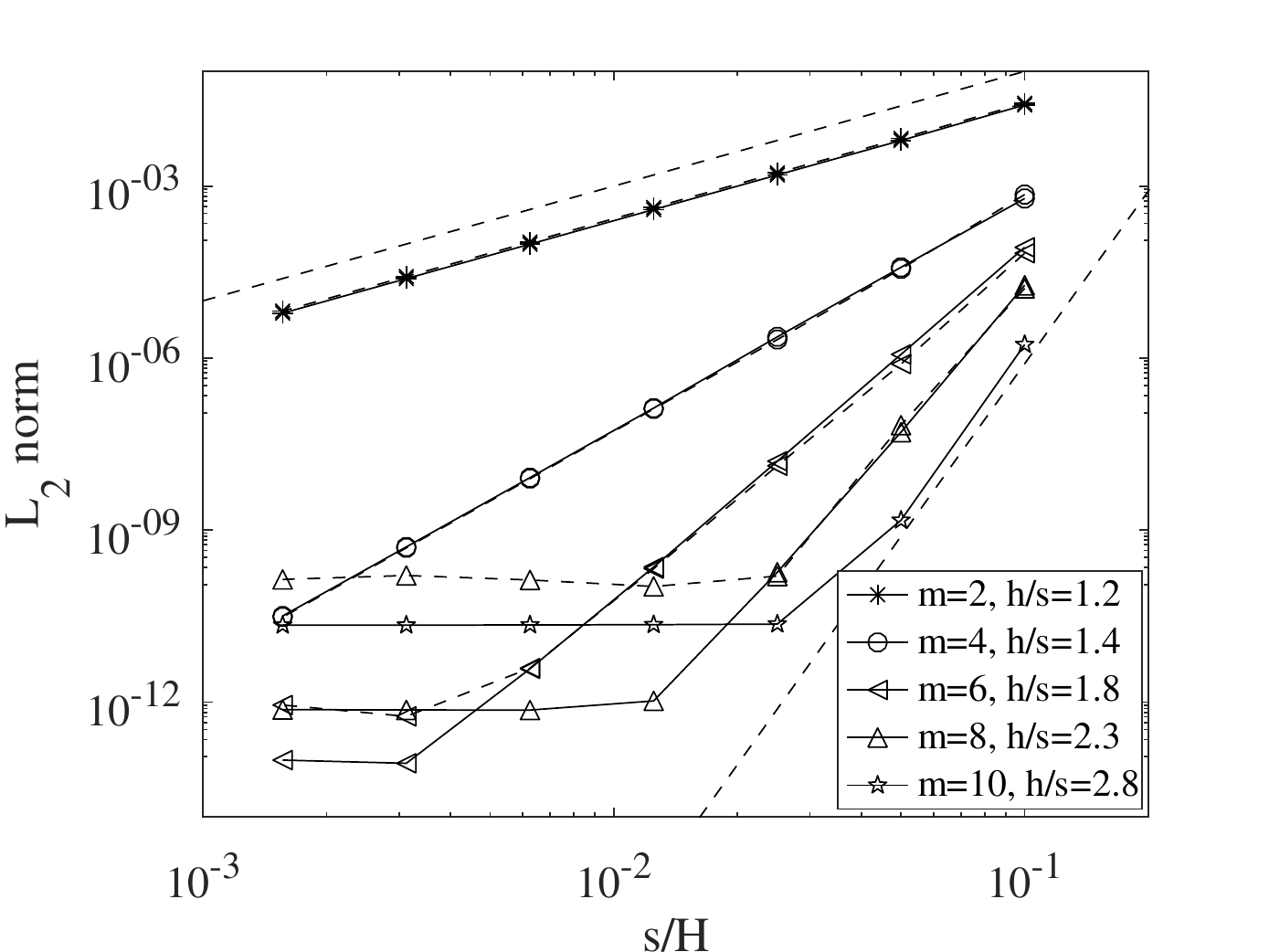}
\includegraphics[width=0.49\textwidth]{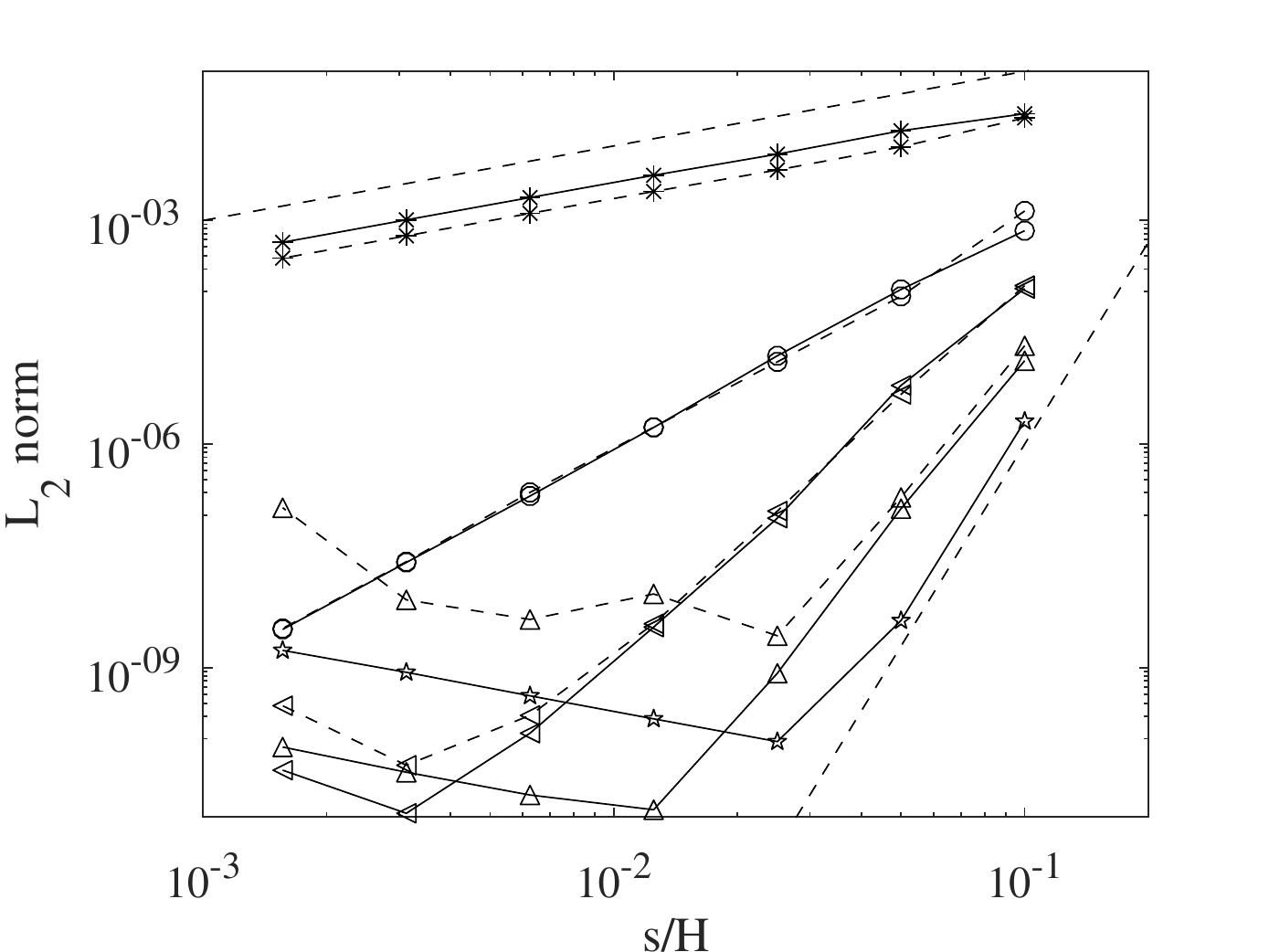}
\caption{Variation of the error in gradient (left) and Laplacian (right) approximations of~\eqref{eq:p6} with resolution $s/H$, for several orders of LABFM. The solid lines are obtained using Hermite ABFs, whilst the dashed lines are obtained using original ABFs.\label{fig:abfconv}}
\end{figure}

Following~\cite{king_2020}, we define the test function
\begin{equation}\phi\left(\hat{x},\hat{y}\right)=1.0+\left(\hat{x}\hat{y}\right)^{4}+\left(\hat{x}\hat{y}\right)^{8}+\displaystyle\sum_{n=1}^{6}\left(\hat{x}^{n}+\hat{y}^{n}\right),\label{eq:p6}\end{equation}
where $\hat{x}=x-0.1453H$ and $\hat{y}=y-0.16401H$, where the translations in $\hat{x}$ and $\hat{y}$ were chosen randomly, to remove any symmetry from $\phi$ which could mask errors in the approximations. We use LABFM to estimate the gradient and Laplacian of $\phi$, over a square domain with side length $H=1$. Figure~\ref{fig:abfconv} shows the convergence of the gradient and Laplacian operator with resolution for a Hermite ABFs and original ABFs, for a range of values of $m$. We set $h/s=\left[1.2,1.4,1.8,2.3,2.8\right]$ for the corresponding order $m$ in the list $m=\left[2,4,6,8,10\right]$. For Hermite ABFs we include the case with $m=10$, whilst for original ABFs we stop at $m=8$.

For small $m$, the results for Hermite ABFs are similar to those for original ABFs. For larger $m$, we see a limiting error (at $\mathcal{O}\left(1\right)$ for gradients and $\mathcal{O}\left(s^{-1}\right)$ for Laplacians) dominate at fine resolutions. For $m=6$ and $m=8$ there is a clear difference between Hermite and original ABFs, with the error limit for Hermite ABFs being more than an order of magnitude smaller. This error limit originates from the accuracy limit of the solution to~\eqref{eq:lsys}. This result is expected: the significantly larger condition number when using original ABFs results in larger errors solving~\eqref{eq:lsys}. It is interesting to note that the error limits we found in~\cite{king_2020} scaled with $s^{-1}$ and $s^{-2}$ for gradients and Laplacians respectively. The difference with the present results arises from the scaling by $\bm{\mathcal{H}}$ described in the previous section: this scaling improves the accuracy of our numerical solution to~\eqref{eq:lsys} by one order. We also see that for increasingly large $m$, the error limit becomes significant. This is a natural consequence of the increasing condition number associated with the larger matrices required for higher order interpolation. There is a compromise to be made in the choice of $m$ - whilst higher values of $m$ incur a greater cost (larger stencils), they generally provide higher accuracy. However, the limiting errors visible in Figure~\ref{fig:abfconv} indicates that larger values of $m$ may provide lower accuracy in over resolved regions, resulting from larger condition numbers. In our simulations of the Navier Stokes equations in Section~\ref{numres}, we choose $m=6$, providing a good compromise of accuracy and computational efficiency.

Insight into the behaviour of the derivative operators can be gained from an analysis of the modal response, following~\cite{brandenburg_2003}. We do this by setting $\phi=\phi\left(\bm{r}_{ji}\right)$ to be a local (to each stencil) sinusoidal function with wavenumber $k$ and, for the derivative under consideration $\bm{C}^{d}\cdot\left.\bm{D}\phi\right\rvert_{i}=k$. For example, to explore the modal response of the convective operator, we set $\phi=\sin{kx_{ji}}$, and for the Laplacian, $\phi=\left(1-\cos{kx_{ji}}\cos{ky_{ji}}\right)/2$. The effective wavenumber is then $k_{eff}=L_{i}^{d}\phi$. The result is normalised with the Nyquist wavenumber (based on the node spacing $s$, $k_{Ny}=\pi/s$). For gradients, the ideal (spectral) response is a linear relation between $k_{eff}$ and $k$, whilst for the Laplacian, the ideal relationship is quadratic. We then take the $L_{2}$ norm of $k_{eff}$ over the domain.

\begin{figure}
\includegraphics[width=0.49\textwidth]{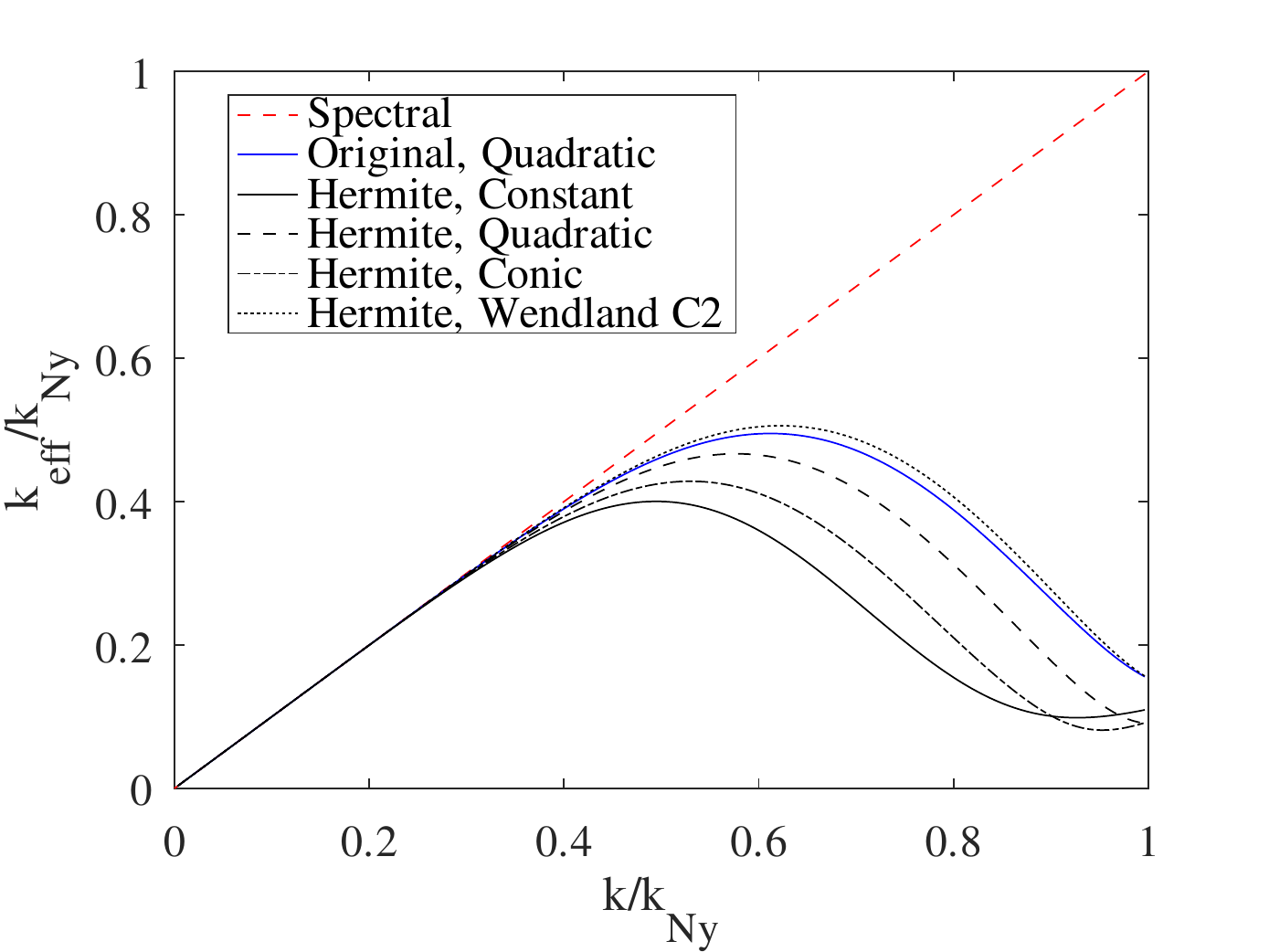}
\includegraphics[width=0.49\textwidth]{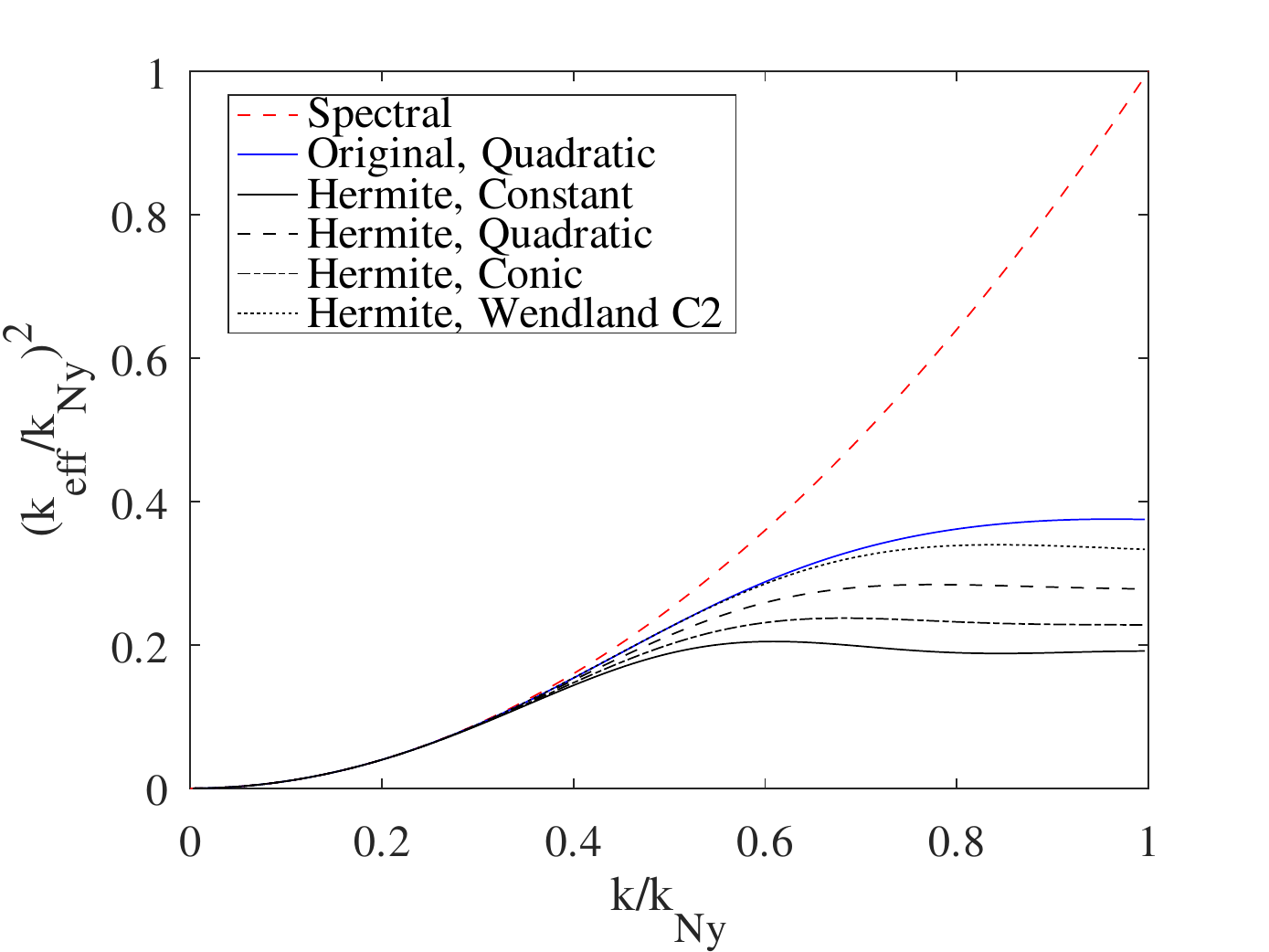}
\caption{The modal response of the gradient (left) and Laplacian (right) operators, for $m=6$, for various choices of ABF and RBF: The `Original' ABF used in~\cite{king_2020} with a quadratic RBF (solid blue lines), and Hermite ABFs (black lines) with a constant RBF (solid line), a quadratic RBF (dashed), a conic RBF (dash-dot), and a Wendland C2 RBF (dotted). The dashed red lines indicate a spectrally accurate method.\label{fig:fr_varABF}}
\end{figure}

Figure~\ref{fig:fr_varABF} shows the modal response for $m=4$, with $h/s=2$, using Hermite ABFs with a range of RBFs, and original ABFs. We see that for higher order RBFs - those with greater structure in the radial direction (i.e. the Wendland RBF) - the modal response improves. We also found (not shown) that for a given ABF and RBF, the modal response improves as $h/s$ is reduced towards its critical value (discussed in detail in~\cite{king_2020}, this is the minimum value of $h/s$, below which the node distribution doesn't adequately sample the ABFs). This suggests that the variation in modal response between RBFs visible in Figure~\ref{fig:fr_varABF} may be due to higher order RBFs having a larger critical value of $h/s$, and the value of $h/s=2$ used here being closer to the limit. This finding suggests that an optimisation procedure could be used to adjust $h/s$ to provide the best modal response. We introduce such a procedure in the next section. Throughout the remainder of this work, we use Hermite ABFs with $\psi$ set to a Wendland C2 RBF.

\section{Stencil optimisation}\label{stencil}

Based on our finding in the previous section, that the modal response of LABFM operators improves for smaller stencils (up to a point), we developed a stencil optimisation procedure. The resulting stencils are both more accurate, and contain fewer neighbours $\mathcal{N}$, reducing computational costs. In an Eulerian context, the optimisation procedure is performed as a preprocessing step at the start of the simulation, prior to calculating the LABFM operator weights. In essence, we reduce $h_{i}$ for every node $i$, until certain criteria are met. The resulting values of $h_{i}$ are close to the minimum (for the given node distribution) stable values, and the LABFM operators yield improved modal responses, relative to non-optimised stencils. The details of the algorithm are as follows. 

All nodes are initialised with $h_{i}=h_{i,0}=2.8s_{i}$. Then, for each node $i$
\begin{enumerate}
\item Set-up and solve the linear system~\eqref{eq:lsys} to calculate the weights for the gradient and Laplacian operators, $\bm{\Psi^{x}}$, $\bm{\Psi^{y}}$, and $\bm{\Psi^{L}}$.
\item Calculate $L_{2}$ norm of the residual of the linear system for the Laplacian operator $\alpha_{L}=\left\lvert\lvert\bm{M}_{i}\bm{\Psi}_{i}-\bm{C^{L}}\right\rvert\rvert_{2}$. If $\alpha_{L}>\alpha_{0}=\frac{10^{-4}\epsilon_{0}n^{4}}{h_{i}m}$, stop. Note that $\epsilon_{0}$ is machine (double) precision zero.
\item For a range of wavenumbers up to the Nyquist wavenumber, $k=\left[k_{Ny}/n_{k},2k_{Ny}/n_{k},\dots,k_{Ny}\right]$, evaluate the amplitude response of the operator. We find that evaluating the amplitude response for $n_{k}=16$ wavenumbers provides reasonable results. We calculate the amplitude response of the operator at wavenumber $k$ using
\begin{subequations}
\begin{align}A^{x}&=\frac{1}{k}L_{i}^{x}\left(\sin\left(kx_{ji}\right)\right)\\A^{y}&=\frac{1}{k}L_{i}^{y}\left(\sin\left(ky_{ji}\right)\right)\\A^{L}&=\frac{1}{k^{2}}L_{i}^{L}\left(0.5-0.5\cos\left(kx_{ji}\right)\cos\left(ky_{ji}\right)\right)\end{align}
\end{subequations}
\item If $\max\left(A^{x},A^{y},A^{L}\right)>1.01$, stop.
\item Otherwise, set $h_{i}=0.99h_{i}$ and return to step 1.
\end{enumerate}

The criteria tested in step $2$ ensures that the linear systems~\eqref{eq:lsys} can be accurately solved. The coefficient $10^{-4}$ is appropriate when using LABFM with Hermite ABFs, but may require adjusting for other ABFs. Furthermore, this criteria will be specific to the method used to solve~\eqref{eq:lsys} (we use the Openblas library~\cite{wang_2013} to solve~\eqref{eq:lsys} via LU decomposition with the \verb+dgesv+ routine). The criteria tested in step $4$ ensures that no modes are amplified by the resulting operators. The algorithm will continue until all nodes have a value of $h_{i}$ such that a reduction of $1\%$ would trigger the criteria in either step $2$ or step $4$. For a disordered node distribution, there can be significant variation in amplitude response from node to node. An apparently ``good'' amplitude response as indicated by the $L_{2}$ norm of $k_{eff}$ can mask the fact that some nodes are unstable - significantly amplifying high wavenumber signals. Therefore, the test of the amplitude response is essential, and ensures local stability of each stencil.

\begin{figure}
\includegraphics[width=0.5\textwidth]{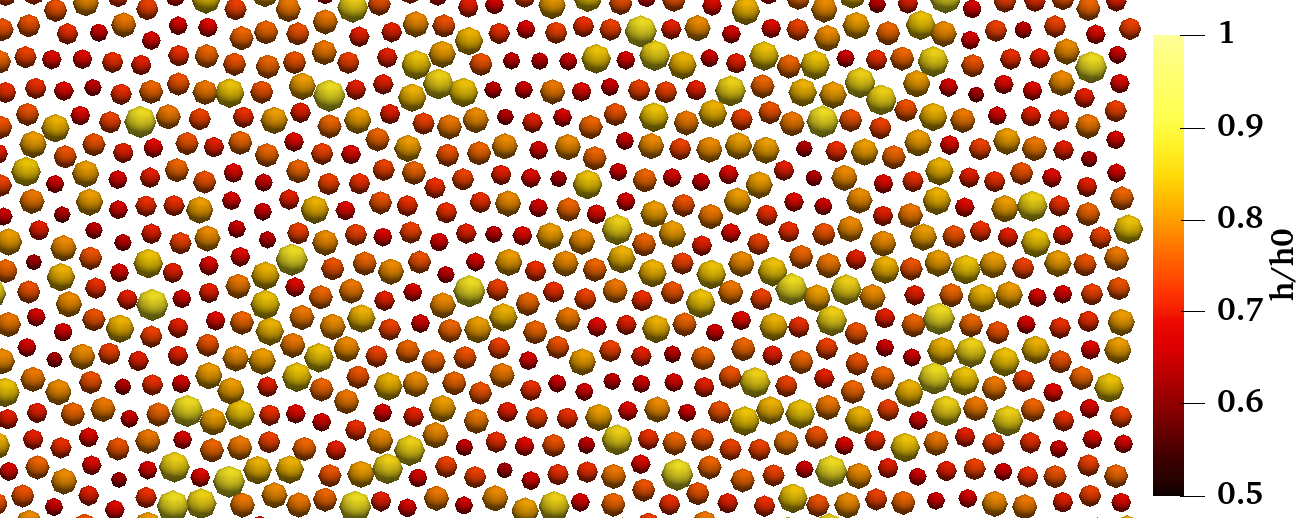}
\caption{A typical node distribution after stencil optimisation. Each circle represents a node, with the size proportional to the number of neighbours $\mathcal{N}_{i}$, and the colour indicating the value of $h_{i}$ (non-dimensionalised by the initial $h_{0,i}$) for that node.\label{fig:stencil_adapt}}
\end{figure}

\begin{figure}
\includegraphics[width=0.49\textwidth]{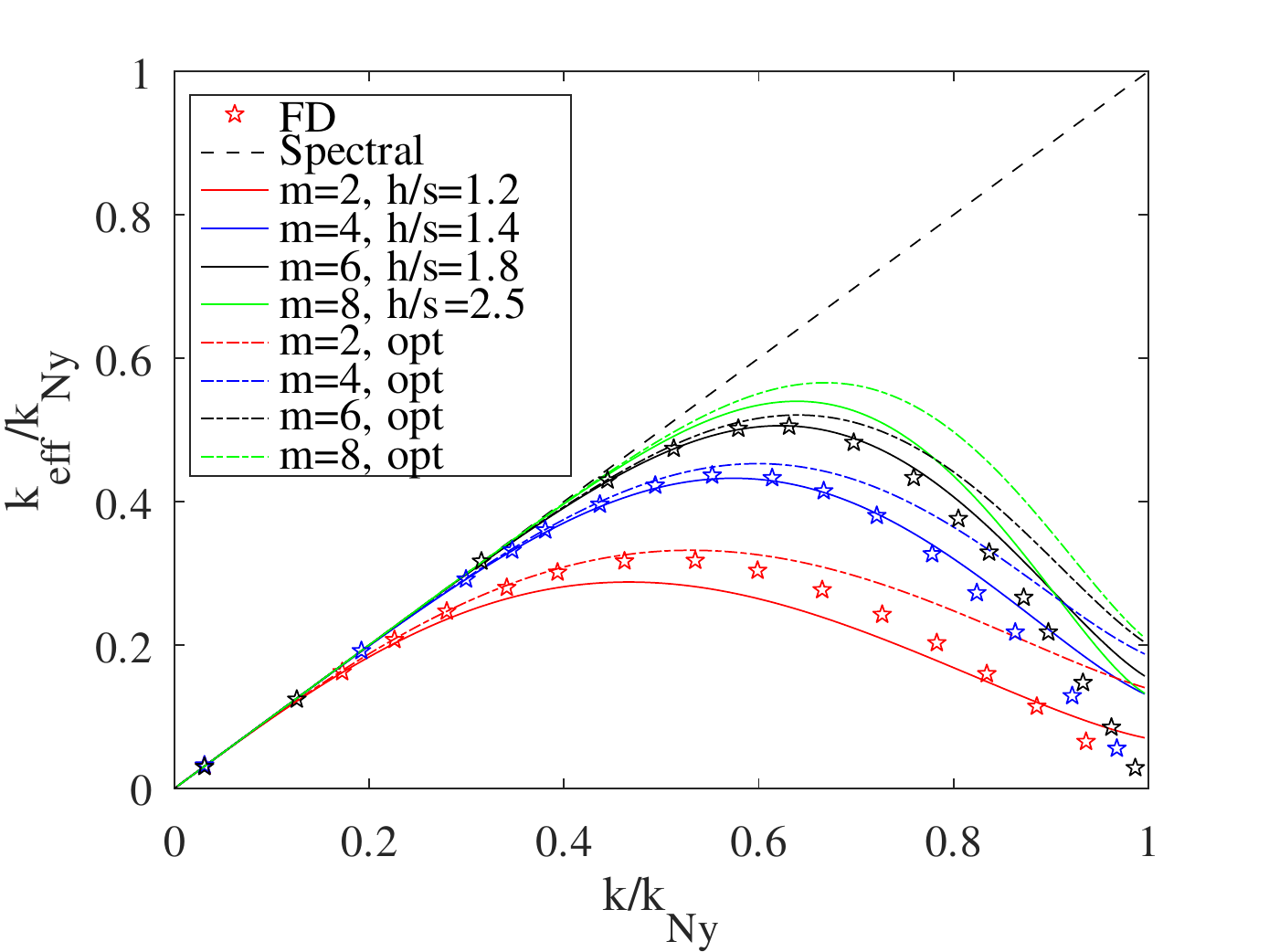}
\includegraphics[width=0.49\textwidth]{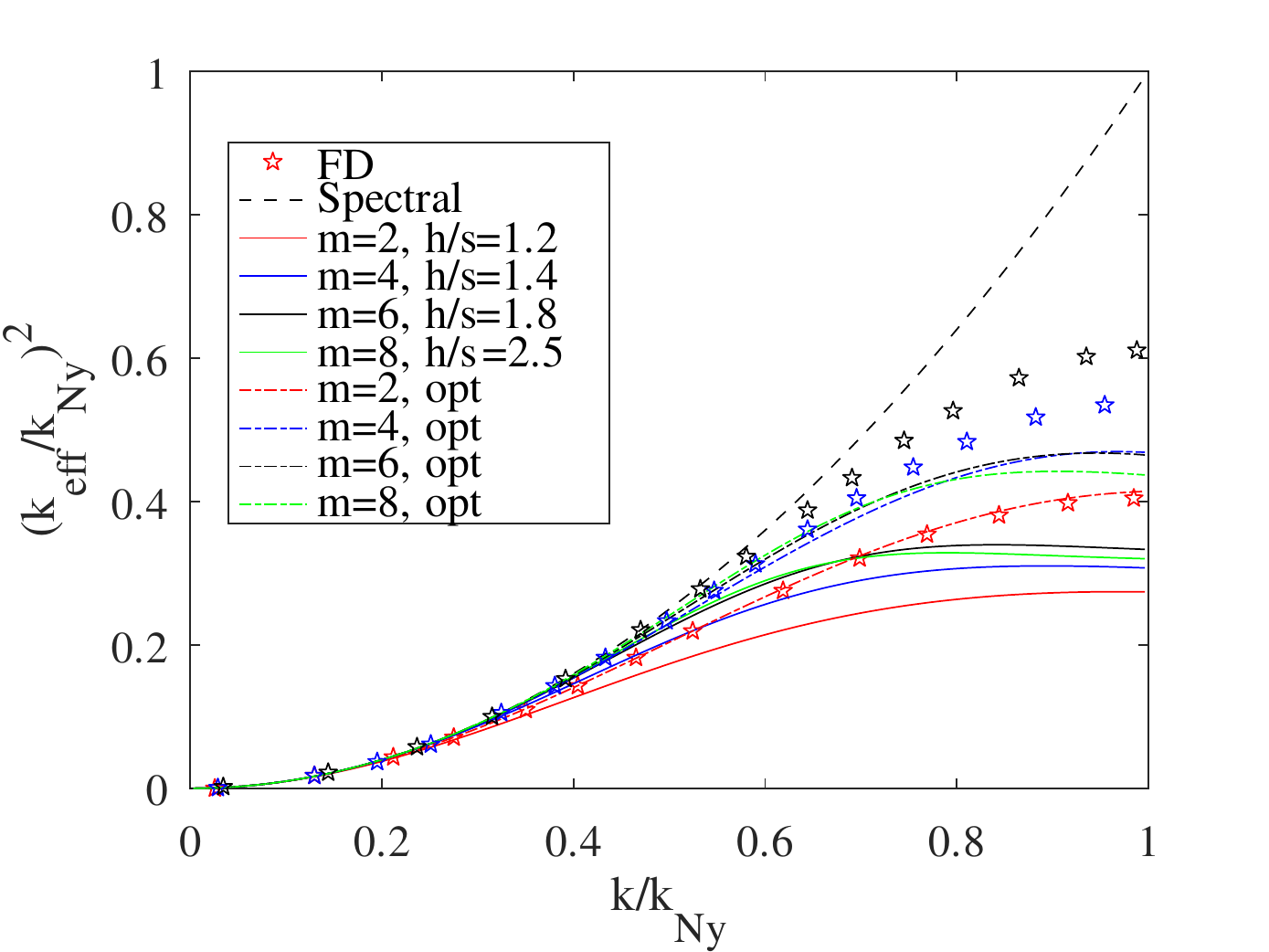}
\caption{The modal response of the gradient (left) and Laplacian (right) operators, for $m=2$ (red lines/symbols), $m=4$ (blue), $m=6$ (black) and $m=8$ (green). The dot-dashed lines indicate the response using our stencil optimisation algorithm, whilst the solid lines are with uniform values of $h/s$. The dashed black lines indicate the ideal modal response, whilst the stars indicate finite differences (from~\cite{brandenburg_2003}).\label{fig:fr_varmopt}}
\end{figure}

Figure~\ref{fig:stencil_adapt} shows a typical node distribution, and the corresponding stencil size $h$ after the optimisation procedure. The size of the circles representing nodes in Figure~\ref{fig:stencil_adapt} is proportional to $h^{2}$, and hence to the number of nodes in the stencil $\mathcal{N}$. There is significant variability in $h$, even though the node spacing $s$ is uniform. Figure~\ref{fig:fr_varmopt} shows the modal response of the gradient and Laplacian operators for $m=2,4,6,8$, both with (dashed lines) and without (solid lines) stencil optimisation. Note that the values of $h/s$ chosen for the case without optimisation are relatively close to the critical $h/s$. The symbols correspond to the modal response for central finite differences, and are taken from~\cite{brandenburg_2003}. For all values of $m$, the optimisation procedure improves the modal response. For gradients, the response with optimised stencils is closer to the ideal case than central finite differences. For Laplacians, finite differences outperform LABFM at high wavenumbers, though the optimisation procedure provides a noticable improvement. Furthermore, the optimisation procedure reduces the computational cost. Taking $m=6$ as an example, the average value of $h/s$ is reduced from $2.2$ (the \emph{safe} non-optimised value to, to approximately $1.5$, reducing the number of neighbours (and hence computational cost of calculating $L_{i}^{d}\phi$) by $17.5\%$. The proportional reduction in stencil size is dependent on the choice of initial stencil size $h_{0}$, and the order of LABFM $m$. We note that there are some nodes for which $h$ is not significantly reduced, and it is likely that this is due to the conservative approach taken in setting our refinement criteria. 
The cost of the optimisation procedure is small compared with numerical simulations of PDEs. We apply the procedure as a preprocessing step, and when using an $8$ core workstation, the algorithm takes approximately $50$ seconds for a distribution with $10^{6}$ nodes. Although this computational cost is dependent on $m$ and $h_{0}$, it is typically of the order of $1$ minute for $10^{6}$ nodes. The algorithm we present here is relatively basic, but demonstrates the potential of stencil optimisation procedures to improve the performance of LABFM (and other consistent mesh-free methods). The development of a more refined optimisation procedure which can give greater improvements in modal response is an active area of research. 

\section{Variable resolution and node generation}\label{varres}

The theoretical basis described in Section~\ref{labfm} is local; the operator $L_{i}^{d}$ is generated based only on a local stencil. In~\cite{king_2020} we limited our investigations to node distributions with uniform resolution: $s_{i}=s_{0}$. However, the formulation does not impose any specific resolution requirements on the global node distribution. All that is required is that the variation of $s_{i}$ is smooth, such that stencils contain enough nodes that the ABFs are adequately sampled. To include spatially varying resolution, we simply need to ensure that $h_{i}$ is adjusted in proportion with $s_{i}$ when generating node distributions. Although largely tangential to the numerical methods detailed herein, we briefly describe our procedure for generating node distributions. For the tests in this work with uniform resolution, we construct a uniform Cartesian distribution (with spacing $s_{i}$), and then shift every node by a random vector $\bm{\varepsilon}=\left(\varepsilon_{x},\varepsilon_{y}\right)^{T}$ (as in~\cite{king_2020}), with the maximum shift set such that $\left\lvert\bm{\varepsilon}\right\rvert/s_{i}\le0.5$. We then apply $10$ iterations of the iterative shifting technique described in~\cite{king_2020}, where nodes a moved according to
\begin{equation}\bm{r}^{n+1}_{i}=\bm{r}^{n}_{i}+\frac{s_{i}^{2}}{h}\displaystyle\sum_{j\in\left\lvert\bm{r}_{ji}\right\rvert<h}\left(\frac{\left\lvert\bm{r}_{ji}\right\rvert}{h}-1\right)\frac{\bm{r}_{ji}}{\left\lvert\bm{r}_{ji}\right\rvert},\label{eq:iter_shift}\end{equation}
with $n$ the iteration number. For cases with spatially varying resolution, we prescribe the resolution as a (smooth) function of position: $s\left(\bm{r}\right)$ (e.g. a function of the form $s\left(\bm{r}\right)=s_{max}-\left(s_{max}-s_{min}\right)e^{-\left\lvert\bm{r}-\bm{r}_{0}\right\rvert^{2}}$ provides increased resolution in the vicinity of $\bm{r}_{0}$). Letting the smallest resolution be $s_{min}$, we then move systematically through the domain in increments of $s_{min}/4$, and place a node if there are no nodes within a distance $s\left(\bm{r}\right)/\sqrt{2}$. Once the entire domain is filled in this manner, we apply $50$ iterations of the shifting procedure described in~\cite{king_2020}. Once the nodes are distributed, we apply the stencil optimisation procedure described in the previous section. An example of the node distribution resulting from this method is shown later in Figure~\ref{fig:cyl_nodes}.

\begin{figure}
\includegraphics[width=0.49\textwidth]{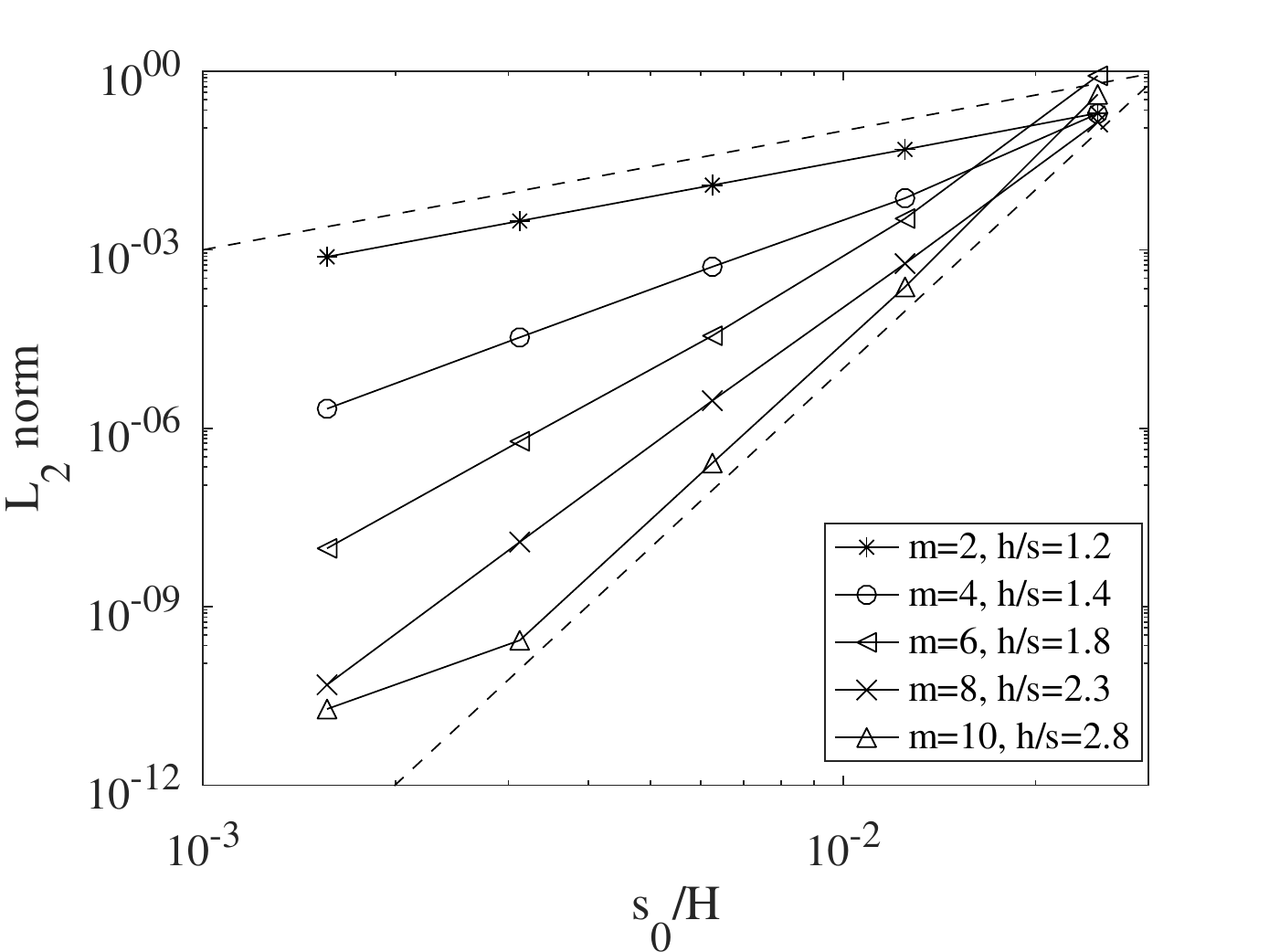}
\includegraphics[width=0.49\textwidth]{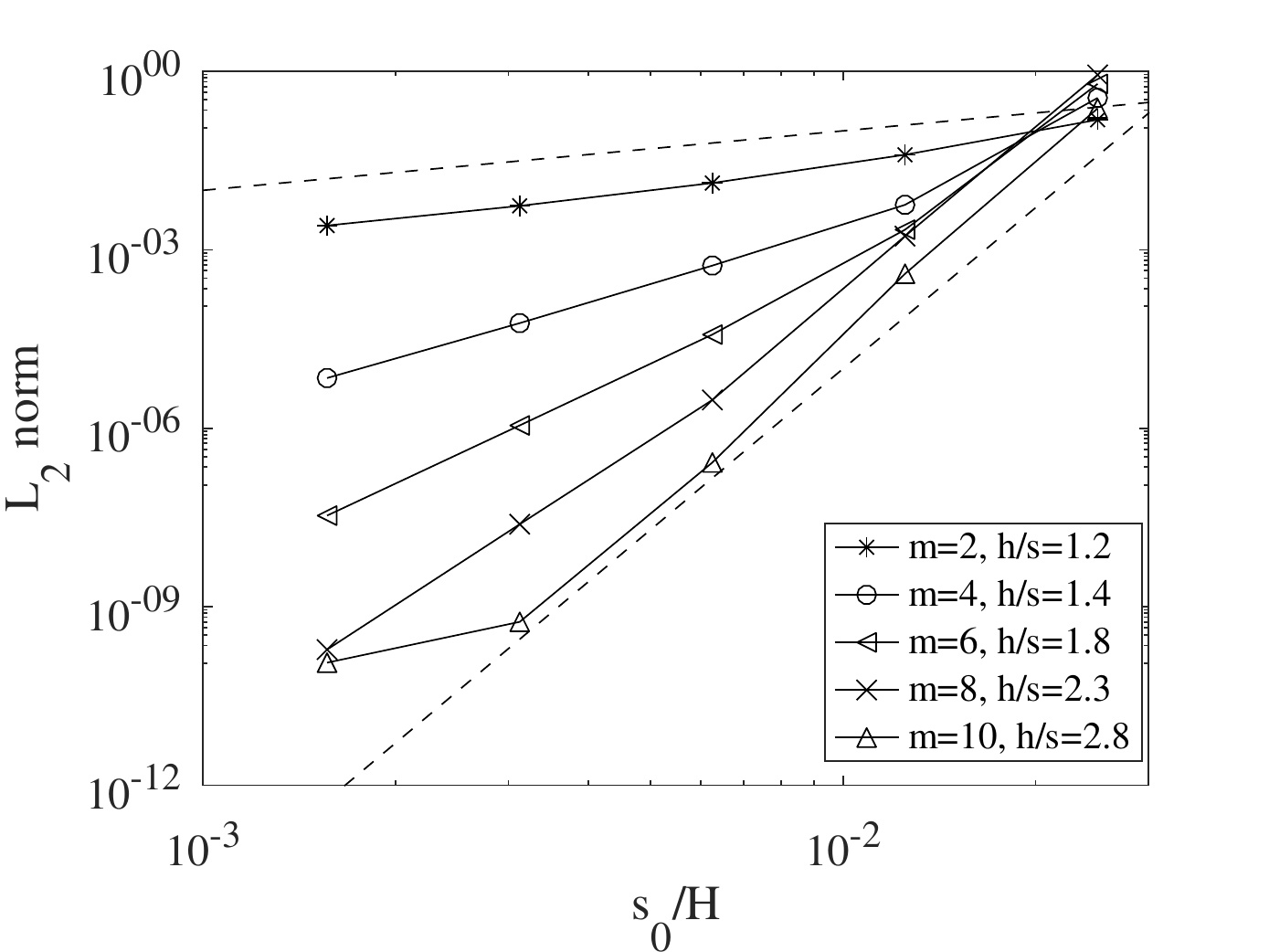}
\caption{Variation of the error in gradient (left) and Laplacian (right) approximations of~\eqref{eq:square} with resolution $s_{0}/H$. The dashed lines show convergence rates of $2$ and $10$ for gradients, and $1$ and $9$ for Laplacians\label{fig:vres}}
\end{figure}

As a brief demonstration of this capability, we test the convergence properties of LABFM on a distribution with non-uniform resolution. The domain is a square with side length $H$, with doubly periodic boundaries. As a test function, we set $\phi$ equal to the first $8$ terms of the Fourier series of a square wave:
\begin{equation}\phi=\sin\left(2\pi{y}/H\right)\frac{4}{\pi}\displaystyle\sum_{k=1}^{8}\frac{\sin\left(2\left(2k-1\right)\pi\left({x}/H-1/4\right)\right)}{2k-1}\label{eq:square}\end{equation}
The function $\phi$ has a steep gradients at $x=0.25$ and $x=0.75$. The node distribution is set such that $s$ is smoothly varying from $s=s_{0}$ at $x=0,1/2,1$ to $s=s_{0}/2$ at $x=0.25,0.75$. The stencil size $h/s$ is spatially uniform, set to $h/s=\left[1.2,1.4,1.8,2.3,2.8\right]$ for $m=\left[2,4,6,8,10\right]$ respectively. Figure~\ref{fig:vres} shows the variation in error with resolution for the gradient (left) and Laplacian (right) approximations for a range of $m$. As with the uniform resolution case, we see convergence rates of $m$ and $m-1$ for the gradient and Laplacian operators respectively.

\section{Spatial filtering}\label{filters}

Collocated discretisation schemes for the Navier Stokes equations admit high wavenumber instabilities. As with high order finite difference schemes on collocated grids, the LABFM operators contain minimal diffusivity, and without special treatment, spurious modes at the Nyquist wavenumber of the node distribution grow and destroy the solution. 
In pseudo-spectral methods, this filtering is performed by de-aliasing of the convective terms~\cite{orszag_1971}. In high order finite difference schemes, spatial filters based on high order derivatives are used~\cite{kennedy_1994,brandenburg_2003}. For RBF based methods, a hyperviscosity term is often added to the governing equations~\cite{fornberg_2011,flyer_2016_ns}, which has a similar effect to explicitly filtering high wavenumbers from the solution. The spatial filters used in finite difference schemes are based on high order partial derivatives, and are typically constructed and applied to the coordinates independently (i.e. a filter based on $\partial^{m}/\partial{x}^{m}$ is applied to rows of nodes aligned with the $x$ coordinate, then one based on $\partial^{m}/\partial{y}^{m}$ is applied to rows aligned with the $y$ coordinate). In contrast, the hyperviscous operators used in~\cite{fornberg_2011} are based on powers of the Laplacian. For example, with $m=4$, the biharmonic operator $\nabla^{4}=\partial^{4}/\partial{x}^{4} + 2\partial^{4}/\partial{x}^{2}\partial{y}^{2}+\partial^{4}/\partial{y}^{4}$ is used. This difference is largely due to the nature of the methods; finite difference stencils are one dimensional, and lend themselves to treating the coordinates separately. Calculation of the hyperviscous operators in RBF-FD methods is straightforwards, and indeed simpler than calculating the individual higher-order partial derivatives. Here LABFM has flexibility to construct a filter based on whichever partial derivatives are required. The stencil is (in two dimensions) circular, and we can readily calculate all the cross-derivatives for the hyperviscous term. We could also, if required, calculated filters for each dimension as in FD schemes. As described in~\cite{king_2020}, we simply modify the values of $\bm{C^{d}}$ to obtain the desired filter. In this work, we construct filters based on hyperviscous terms, as in RBF methods. This approach is consistent with the circular (in two dimensions) stencil used, which may admit high wavenumber instablities which are not necessarily aligned with the coordinate system. Rather than modify the governing equations with additional hyperviscous terms, we explicitly filter the evolved variables after every time step, with the filter coefficient set locally. In particular, such that a dealiasing rule is satisfied on a node-by-node basis.

After each complete time step, evolved variables are filtered according to
\begin{equation}\tilde\phi_{i}=\left(1+\mathcal{F}_{i}\right)\phi,\end{equation}
where the filter operator is $\mathcal{F}_{i}=\kappa_{m,i}\nabla^{m}$, with $m$ (the order of LABFM used) assumed to be even. Note that if we chose an odd value of $m$, the filter would be constructed using $\nabla^{m-1}$. In~\cite{kennedy_1994,brandenburg_2003} the filter coeffient $\kappa_{m}$ is set uniformly, based on the modal response of the the continuous operator $\nabla^{m}$, although this is not the optimum approach. In fact, the wavenumber response of the filter is likely to deviate significantly from the ideal case, resulting in under-damping at high wavenumbers. For our purposes, there is also likely to be variation in the response of the filter between nodes, due to the disorder of the node distribution. We introduce here a technique for measuring the response of the filter and adjusting the coefficient $\kappa_{m,i}$ accordingly. In a preprocessing step, we calculate
\begin{equation}F_{0,i}=\displaystyle\sum_{j}\phi_{ji}w_{ji}^{hyp},\qquad\text{with}\qquad\phi_{ji}=1-\cos\left(\frac{3\pi{x}_{ji}}{2s_{i}}\right)\cos\left(\frac{3\pi{y}_{ji}}{2s_{i}}\right),\end{equation}
in which $w_{ji}^{hyp}$ are the LABFM-generated weights for the operator $\nabla^{m}$. We then set
\begin{equation}\kappa_{m,i}=\frac{2}{3F_{0,i}}.\end{equation}
This procedure ensures that the amplitude response of the filter at wavelengths two-thirds of the Nyquist wavelength is $1/3$.
With this approach it is easy to vary the dealiasing rule: we have also investigated setting the amplitude response to zero at two-thirds the Nyquist wavelength, or to zero at the Nyquist wavelength. We found that setting the amplitude response at two-thirds the Nyquist wavenumber to $1/3$ provided the most reliable stabilising effect, accounting for the variability in filter response between nodes due to the anisotropy of the node distribution. The value of $\kappa_{m}$ obtained with this approach is proportional to $s^{m}$, whilst the accuracy of the discrete operator approximating $\nabla^{m}$ is first order. Therefore the resulting filters introduce a leading error of $\mathcal{O}\left(s^{m+1}\right)$. This error is of higher order than the discretisation scheme for spatial derivatives (which is order $m$) and hence the overall order the method is not affected.

This filtering regime enables stable numerical simulations of the compressible Navier-Stokes equations, as demonstrated in the following sections. We note that although in this work our focus is on compressible flows, the addition of this filtering technique to an implementation of LABFM in a fractional step algorithm to solve the incompressible Navier-Stokes equations provides stable results. The application of LABFM to simulate the incompressible Navier-Stokes equations will be the subject of a separate work.

\section{Implementation for simulations of isothermal flows}\label{ns}

We consider the compressible Navier Stokes equations under the assumption of isothermal flow. The Navier Stokes equations can be expressed in terms of density-logarithm $\ln\rho$ and velocity $\bm{u}$ as
\begin{subequations}\begin{align}\frac{\partial\ln\rho}{\partial{t}}+\bm{u}\cdot\nabla\ln\rho&=-\nabla\cdot\bm{u},\label{eq:mass}\\\frac{\partial\bm{u}}{\partial{t}}+\bm{u}\cdot\nabla\bm{u}&=\frac{-\nabla{p}}{\rho}+\frac{1}{\rho}\nabla\cdot\bm{\tau}+\bm{g},\label{eq:mom}\end{align}\end{subequations}
in which $\bm{g}$ is a body force (e.g. due to gravity) and $p$ is the pressure. Unless explicitly stated otherwise, we set $\bm{g}=\bm{0}$. The viscous stress is
\begin{equation}\bm{\tau}=\mu\left(\nabla\bm{u}+\nabla\bm{u}^{T}-\frac{2}{3}\bm{I}\nabla\cdot\bm{u}\right),\end{equation}
where the viscosity $\mu$ is assumed to be constant and uniform, bulk viscosity is neglected, and $\bm{I}$ is the identity tensor. Under the isothermal assumption, we can use a polytropic equation of state with an exponent of unity to close the system:
\begin{equation}\frac{p}{\rho}=c^{2}\end{equation}
in which $c^{2}$ is the (constant and uniform) speed of sound. In all cases we consider subsonic flows. For two-dimensional flows we have $\bm{u}=\left(u,v\right)^{T}$, and the system~\eqref{eq:mass} and~\eqref{eq:mom} may be written
\begin{subequations}\begin{align}
\frac{\partial\ln\rho}{\partial{t}}+u\frac{\partial\ln\rho}{\partial{x}}+v\frac{\partial\ln\rho}{\partial{y}}&=-\frac{\partial{u}}{\partial{x}}-\frac{\partial{v}}{\partial{y}}\\
\frac{\partial{u}}{\partial{t}}+u\frac{\partial{u}}{\partial{x}}+v\frac{\partial{u}}{\partial{y}}&=\frac{-1}{\rho}\frac{\partial{p}}{\partial{x}}+\frac{\mu}{\rho}\left[\frac{4}{3}\frac{\partial^{2}{u}}{\partial{x}^{2}}+\frac{\partial^{2}{u}}{\partial{y}^{2}}+\frac{1}{3}\frac{\partial^{2}v}{\partial{x}\partial{y}}\right]+g_{x}\\
\frac{\partial{v}}{\partial{t}}+u\frac{\partial{v}}{\partial{x}}+v\frac{\partial{v}}{\partial{y}}&=\frac{-1}{\rho}\frac{\partial{p}}{\partial{y}}+\frac{\mu}{\rho}\left[\frac{4}{3}\frac{\partial^{2}{v}}{\partial{y}^{2}}+\frac{\partial^{2}{v}}{\partial{x}^{2}}+\frac{1}{3}\frac{\partial^{2}u}{\partial{x}\partial{y}}\right]+g_{y}.\end{align}\label{eq:icns}\end{subequations}
We evaluate spatial derivatives using LABFM with and Hermite ABFs (with a Wendland C2 RBF as defined in~\cite{king_2020}). We find that setting $m=6$ provides a good compromise between computational efficiency, accuracy, and stability. In the following, if not explicitly specified otherwise, we set $m=6$ in the bulk of the domain, whilst near boundaries, we reduce the order of LABFM to $m=4$ as described in below in Section~\ref{boundaries}. We integrate~\eqref{eq:icns} in time using a four-step third-order, low-storage explicit Runge-Kutta denoted RK3(2)4[2R+]C in the classification system of~\citet{kennedy_2000}. The value of the time-step $\delta{t}$ set by the CFL and viscous criteria
\begin{equation}\delta{t}=\min_{\Omega}\left\{\min\left[1.0\frac{s}{\left\lvert\bm{u}\right\rvert+c},0.4\frac{s^{2}\rho}{\mu}\right]\right\},\label{eq:dt}\end{equation}
where $\displaystyle\min_{\Omega}$ is the minimum over the entire domain. We apply the filter described in Section~\ref{filters} to the evolved variables $\ln\rho$, $u$ and $v$ after each complete time step. In practice, for all flows considered herein, the convective constraint (the first term in~\eqref{eq:dt}) is the most stringent.

\subsection{Boundaries}\label{boundaries}

On boundaries, where the computational stencil has only partial support, the difference operators generated by LABFM do not always converge, as the linear systems $\bm{M}_{i}$ become poorly conditioned. In~\cite{king_2020} we found that for one-sided LABFM (that is, where half of the domain of support of a node contains no neighbouring nodes e.g. node $i_{0}$ in Figure~\ref{fig:stencils}), the method converged for $m\le4$, although with significantly increased error magnitude, whilst for $m>4$, the conditioning of the linear systems $\bm{M}_{i}$ was such that the solver didn't yield a valid solution. A new stability analysis since~\cite{king_2020} (via the eigenvalues of the global derivative operator) has shown that for one-sided LABFM with $m=4$ the eigenvalues spread along the positive real axis: the operators are unstable. For nodes with nearly one-sided stencils (e.g. node $i_{1}$ in Figure~\ref{fig:stencils}), the viscous operators are stable, whilst the convective operators remain unstable, although less so than the fully one-sided nodes. A one-sided variant of LABFM which yields high accuracy whilst retaining the stability properties of LABFM at internal nodes is extremely desirable, both for the subject of this section - boundaries - and for the potential to develop an upwind scheme. Such a developement is an area of research we are actively pursuing. 

In this work, we bypass the issue of one-sided LABFM, by using a combination of LABFM and finite differences near domain boundaries. To achieve a $4^{th}$ order discretisation scheme on boundaries, we require a strip within the fluid that is $5$  uniformly distributed nodes thick along the boundaries. In practice, this doesn't impose a particularly severe restriction on the complexity of the geometries we can model. This $5$ node thick, locally orthogonal distribution can follow a boundary of arbitrarily complex shape, provided the distribution lengthscale is adapted to resolve the boundary curvature. The present discretisation scheme requires minimal information to describe the boundaries: an unordered set of boundary nodes, each with a position and a unit normal vector. In the present work we focus on flows through multiply connected domains, for which we require inflow, outflow, and wall boundary conditions, alongside (trivial) periodic or symmetry conditions. In the following we first describe our discretisation strategy, and then our method for implementing boundary conditions.

\subsubsection{Boundary discretisation} 

\begin{figure}
\includegraphics[width=0.32\textwidth]{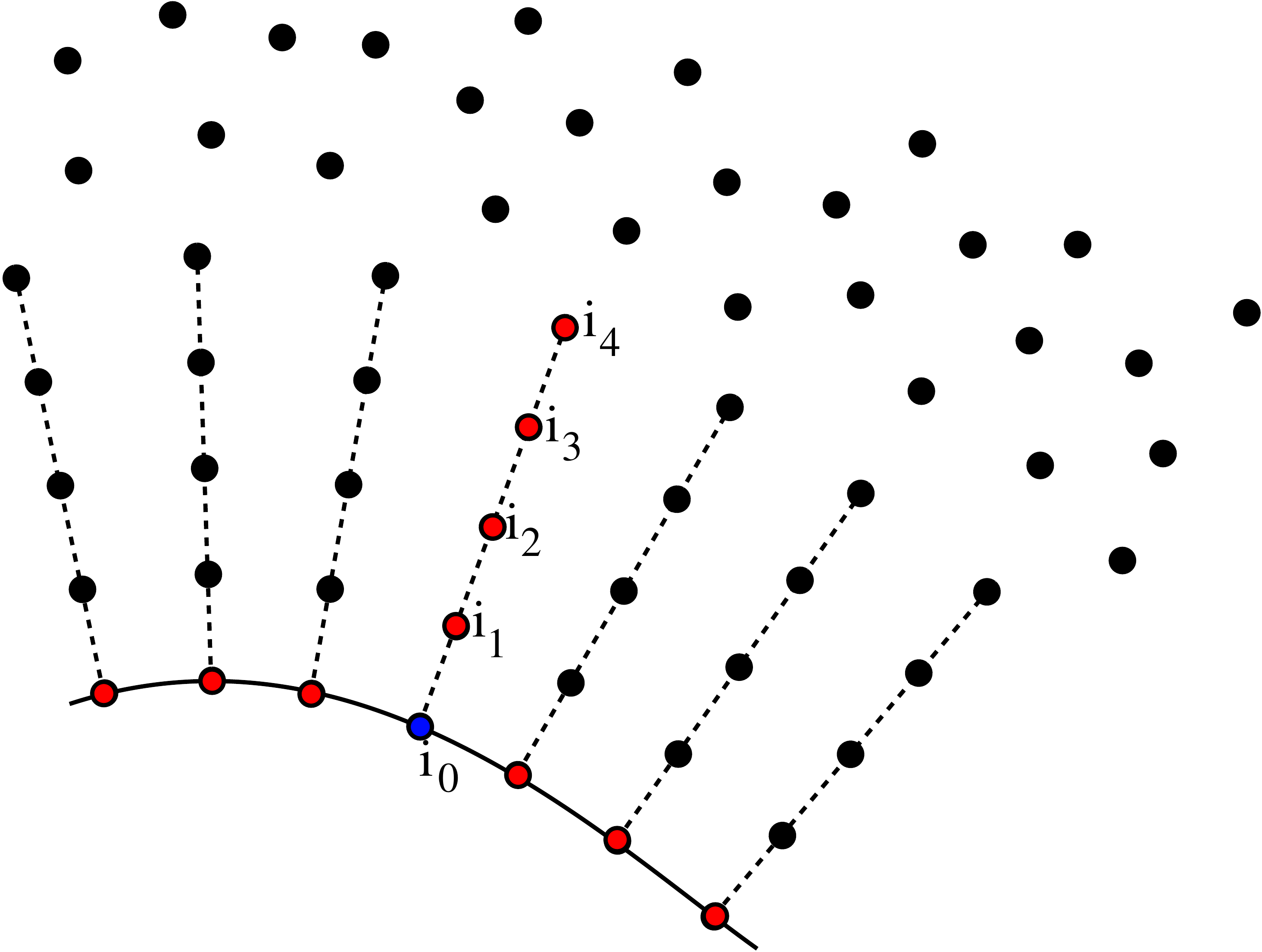}
\includegraphics[width=0.32\textwidth]{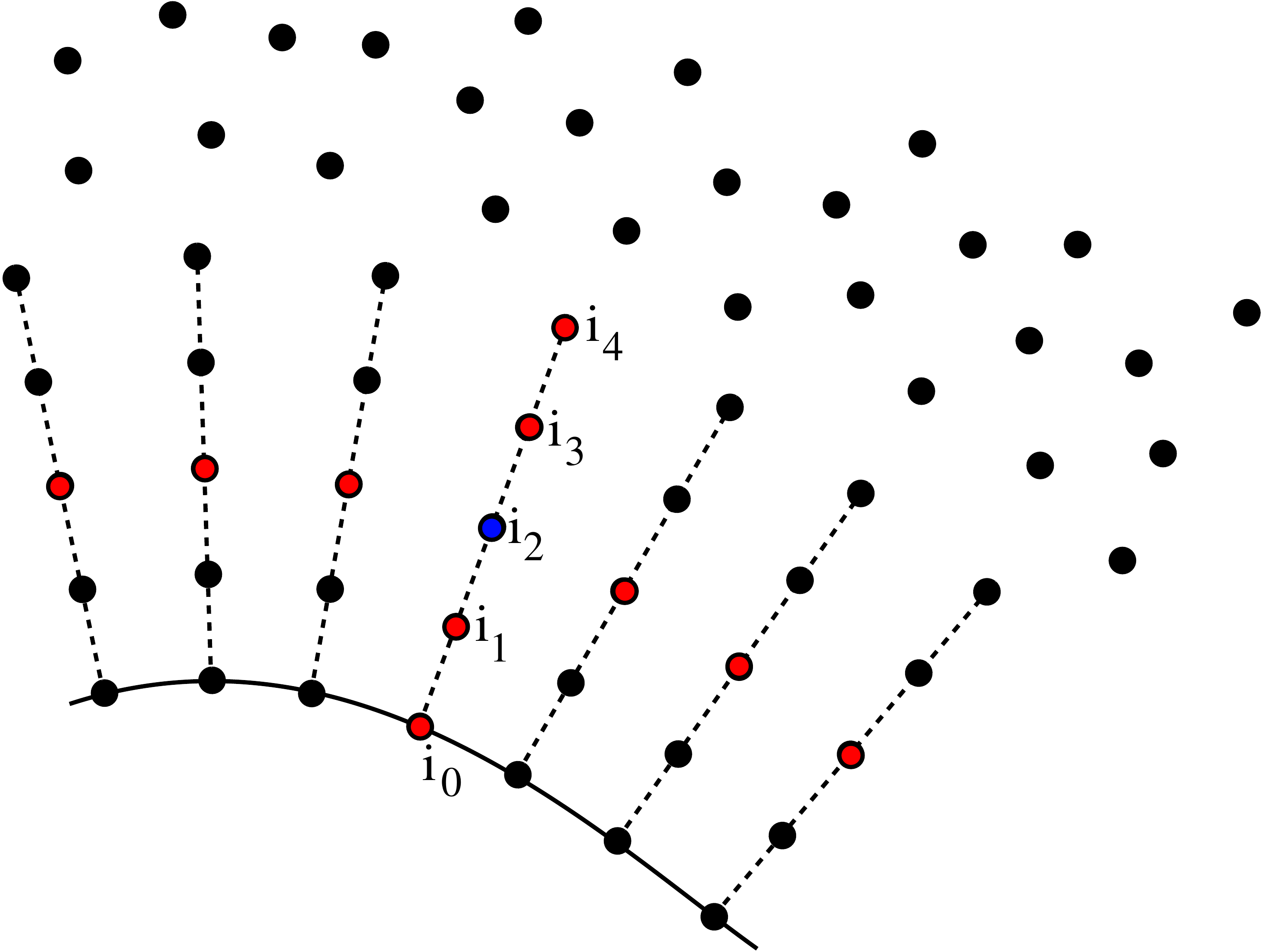}
\includegraphics[width=0.32\textwidth]{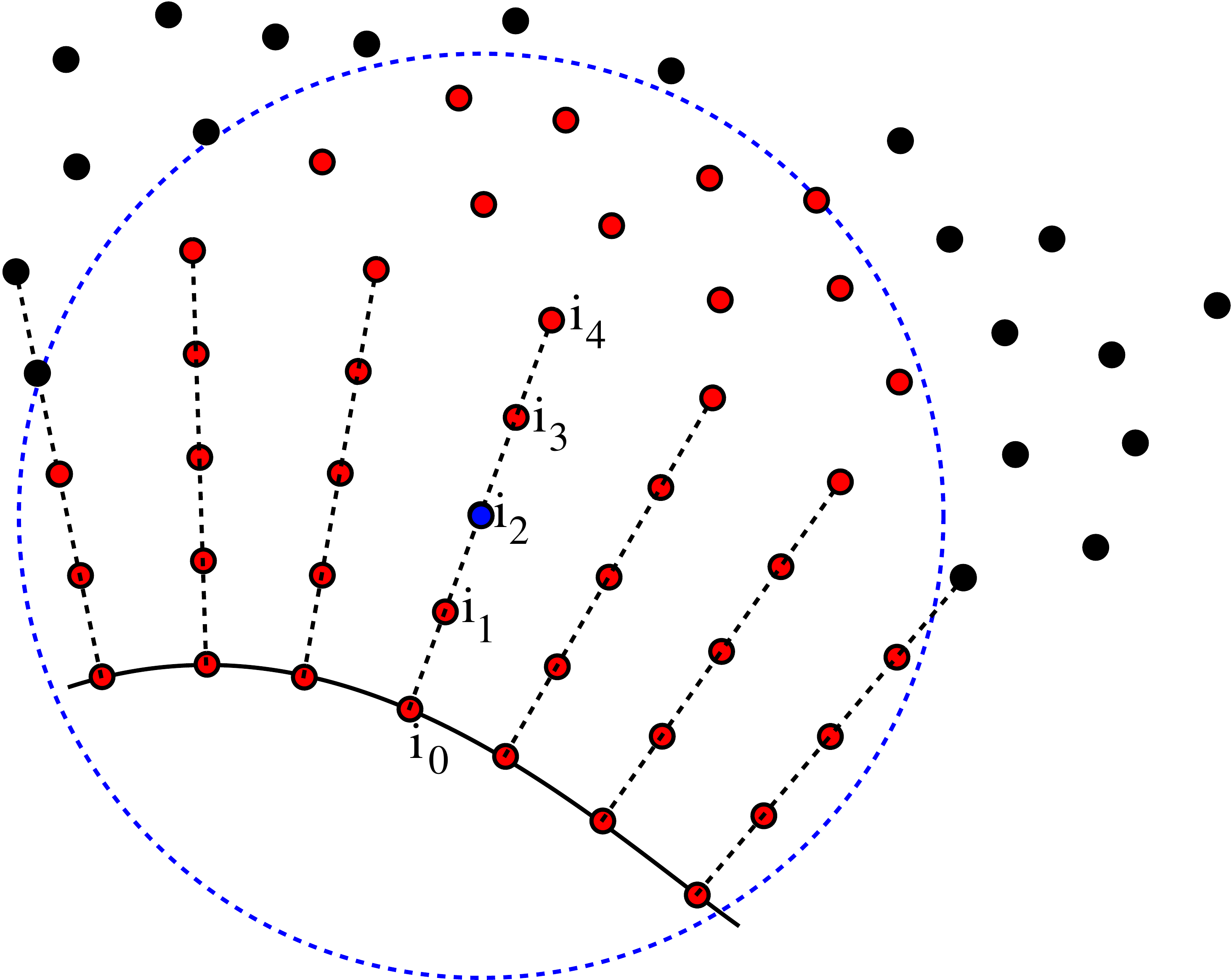}
\caption{The computational stencils used near domain boundaries. The boundary is represented by a solid black line, and the locally orthogonal node distribution shown along boundary normals (dashed-lines). The left panel shows the stencil for a boundary node (denoted $i_{0}$). The centre panel shows the stencil for first derivatives of a near-boundary node (denoted $i_{2}$). The right panel shows the stencil for the second and higher order derivatives at a near boundary node. In all cases, the nodes within the stencil are highlighted in red, and the node at which the derivatives are to be calculated is highlighted in blue.\label{fig:stencils}}
\end{figure}

Boundaries and nearby fluid are discretised as follows, and as illustrated in Figure~\ref{fig:stencils}. The set of nodes on the boundary is denoted $B$, and for every node $i_{0}\in{B}$, the unit normal vector (pointing into the domain) is $\bm{n}_{i_{0}}$, and a tangent vector $\bm{t}_{i_{0}}$. We denote the coordinate system aligned with the (local) boundary normal and tangets as $\left(\xi,\eta\right)$, and the normal and tangential components of velocity as $u_{\xi}$ and $u_{\eta}$ respectively. The node distribution is initialised such that every boundary node $i_{0}$ has neighbours at $\bm{r}_{i_{0}}+qs_{i_{0}}\bm{n}_{i_{0}}$ for $q=1,2,3,4$ (these nodes are denoted here as $i_{1}$, $i_{2}$, $i_{3}$ and $i_{4}$). On nodes $i_{0}\in{B}$, boundary normal derivatives are then calculating using $5$ point one-sided finite differences. Boundary tangent derivatives are obtained using only nodes lying on the boundary, with a one-dimensional formulation of LABFM, detailed in Appendix~\ref{1dlabfm}. These stencils are illustrated in the left panel of Figure~\ref{fig:stencils}. For nodes $i_{1}$ and $i_{2}$, asymmetric and central $5$ point finite differences are used respectively to evaluate boundary normal derivatives, whilst boundary transverse derivatives are calculated with one-dimensional LABFM. For nodes $i_{1}$ and $i_{2}$, second derivatives are evaluated in the $\left(x,y\right)$ coordinate system using two-dimensional LABFM. The convective and viscous stencils for node $i_{2}$ are illustrated in the central and right panels (respectively) of Figure~\ref{fig:stencils}. For nodes $i_{1}$ and $i_{2}$, the first derivatives in the $\left(x,y\right)$ coordinate system are obtained by multiplying by the Jacobian
\begin{equation}\begin{bmatrix}\frac{\partial\phi}{\partial{x}}\\\frac{\partial\phi}{\partial{y}}\end{bmatrix}=\begin{bmatrix}\bm{n}_{i_{0}}\cdot\bm{e_{x}}&-\bm{n}_{i_{0}}\cdot\bm{e_{y}}\\\bm{n}_{i_{0}}\cdot\bm{e_{y}}&\bm{n}_{i_{0}}\cdot\bm{e_{x}}\end{bmatrix}\begin{bmatrix}\frac{\partial\phi}{\partial\xi}\\\frac{\partial\phi}{\partial\eta}\end{bmatrix}.\end{equation}
Whilst for the internal scheme, we impose no limit on the order of LABFM $m$, on nodes $i_{1}$ and $i_{2}$, the stencils are incomplete, and LABFM locally fails to converge for $m>4$. Therefore, on nodes $i_{1}$ and $i_{2}$ second (and higher) derivatives are calculated using $4^{th}$ order LABFM. For nodes $i_{3}$, $i_{4}$, and all other internal nodes, all derivatives are obtained via LABFM as described in the previous sections. Table~\ref{tab:boundaries} details the spatial discretisation schemes for nodes $i_{0}$ to $i_{4}$.

\begin{table}
\begin{center}
\caption{Schemes for derivatives near boundaries.\label{tab:boundaries}}
\begin{tabular}{|l|c|c|c|c||}
\hline
\textbf{Node} & $\partial\phi/\partial\xi$ & $\partial\phi/\partial\eta$ & Second derivatives & Hyperviscous operator \\
\hline
$i_{0}$ & One-sided FD & 1D-LABFM & FD + 1D-LABFM & FD + 1D-LABFM \\
$i_{1}$ & Assymetric FD & 1D-LABFM & LABFM ($m=4$) & LABFM ($m=4$) \\
$i_{2}$ & Centered FD & 1D-LABFM & LABFM ($m=4$) & LABFM ($m=4$) \\
$i_{3}$ & LABFM & LABFM & LABFM & LABFM  \\
$i_{4}$ & LABFM & LABFM & LABFM & LABFM  \\
\hline
\end{tabular}
\end{center}
\end{table}

\subsubsection{Boundary conditions}\label{bound}

With the above discretisation scheme in place, boundary conditions are imposed using the Navier-Stokes characteristic boundary condition (NSCBC) formulation. Initially developed for inviscid flows~\cite{thompson_1987,thompson_1990}, and subsequently extended to the Navier-Stokes equations~\cite{poinsot_1992,sutherland_2003}, the method involves a Locally One-Dimensional Inviscid (LODI) assumption, allowing the boundary normal convective terms to be treated separately from the tangential, viscous and source terms. The boundary normal convective terms are formulated in terms of characteristic wave amplitudes, and waves which are incoming at the boundary are specified to enforce the appropriate physical boundary conditions. The system~\eqref{eq:icns} is written in the boundary-oriented coordinate system $\left(\xi,\eta\right)$, and then following~\citet{sutherland_2003}, the convective terms normal to the boundary are expressed in terms of characteristic wave amplitudes as
\begin{subequations}\begin{align}
\frac{\partial\ln\rho}{\partial{t}}+\frac{1}{\rho{c}^{2}}\left(c^{2}\mathcal{L}_{2}+\mathcal{L}_{4}+\mathcal{L}_{1}\right)+u_{\eta}\frac{\partial\ln\rho}{\partial\eta}&=-\frac{\partial{u}_{\eta}}{\partial\eta},\label{eq:lodilnro}\\
\frac{\partial{u}_{\xi}}{\partial{t}}+\frac{1}{\rho{c}}\left(\mathcal{L}_{4}-\mathcal{L}_{1}\right)+u_{\eta}\frac{\partial{u_{\xi}}}{\partial\eta}&=\frac{\mu}{\rho}\left[\frac{4}{3}\frac{\partial^{2}{u}_{\xi}}{\partial{\xi}^{2}}+\frac{\partial^{2}{u}_{\xi}}{\partial{\eta}^{2}}+\frac{1}{3}\frac{\partial^{2}{u}_{\eta}}{\partial{\xi}\partial{\eta}}\right]+g_{\xi},\label{eq:lodiu}\\
\frac{\partial{u}_{\eta}}{\partial{t}}+\mathcal{L}_{3}+u_{\eta}\frac{\partial{u}_{\eta}}{\partial\eta}&=\frac{-1}{\rho}\frac{\partial{p}}{\partial{\eta}}+\frac{\mu}{\rho}\left[\frac{4}{3}\frac{\partial^{2}{u}_{\eta}}{\partial{\eta}^{2}}+\frac{\partial^{2}{{u}_{\eta}}}{\partial{\xi}^{2}}+\frac{1}{3}\frac{\partial^{2}{u}_{\xi}}{\partial{\xi}\partial{\eta}}\right]+g_{\eta}\label{eq:lodiv}.\end{align}\label{eq:lodi}\end{subequations}
The symbols $\mathcal{L}_{1}$ to $\mathcal{L}_{4}$ are the characteristic wave amplitudes describing the propagation of acoustic, entropy, vorticity and acoustic waves respectively, and are defined by:
\begin{subequations}
\begin{align}
\mathcal{L}_{1}&=\frac{u_{\xi}-c}{2}\left(\frac{\partial{p}}{\partial\xi}-\rho{c}\frac{\partial{u}_{\xi}}{\partial\xi}\right)\\
\mathcal{L}_{2}&=\frac{u_{\xi}}{c^{2}}\left(c^{2}\frac{\partial\rho}{\partial\xi}-\frac{\partial{p}}{\partial\xi}\right)\\
\mathcal{L}_{3}&=u_{\xi}\frac{\partial{u}_{\eta}}{\partial\xi}\\
\mathcal{L}_{4}&=\frac{u_{\xi}+c}{2}\left(\frac{\partial{p}}{\partial\xi}+\rho{c}\frac{\partial{u}_{\xi}}{\partial\xi}\right).
\end{align}\label{eq:Ldef}
\end{subequations}
Note that for the isothermal flow assumed here, entropy is constant, and $\mathcal{L}_{2}=0$ by definition. The wave speeds associated with the quantities $\mathcal{L}_{1}$ to $\mathcal{L}_{4}$ are $u_{\xi}-c$, $u_{\xi}$, $u_{\xi}$ and $u_{\xi}+c$ in the direction of $\bm{n}_{i}$. In this work we consider only stationary boundaries. On any boundary, those $\mathcal{L}$ which are propagating out of the domain should be calculated from~\eqref{eq:Ldef} using information within the domain. Those $\mathcal{L}$ which are propagating into the domain must be specified as numerical boundary conditions as described below. In certain cases, additional boundary conditions are imposed on the viscous terms. Whilst we solve~\eqref{eq:icns} within the domain, on the boundaries we solve~\eqref{eq:lodi}. It is known that the original NSCBC formulation encounters difficulties when vorticity passes through outflow boundaries~\cite{fosso_2012}. To overcome this issue, we largely follow the modifications of~\cite{yoo_2007}, which improve the ability of the boundary conditions to handle vorticity by acounting for transverse and viscous terms in the specification of incoming wave amplitudes.

At a no-slip wall, we have $\bm{u}=0$, so $\mathcal{L}_{4}$ is propagating into the domain and must be specified. We do this by setting $u_{\xi}=u_{\eta}=0$, and updating $\ln\rho$ with~\eqref{eq:lodilnro} using
\begin{equation}\mathcal{L}_{4}=\mathcal{L}_{1}+2\rho{c}\bm{g}\cdot\bm{n}_{i}.\end{equation}
This enforces the velocity on the wall to be zero, and results in perfect reflection of acoustic waves. 

On outflow boundaries $\mathcal{L}_{1}$ to $\mathcal{L}_{3}$ are outgoing, and $\mathcal{L}_{4}$ is incoming (in the boundary-oriented coordinate system). Hence $\mathcal{L}_{1}$ to $\mathcal{L}_{3}$ are calculated from their definitions in~\eqref{eq:Ldef}, whilst $\mathcal{L}_{4}$ is specified to ensure the boundary is transparent to acoustic waves and vorticity. To achieve this, we follow the approach of~\citet{yoo_2007}, setting
\begin{equation}\mathcal{L}_{4}=\frac{0.278}{2L}c\left(1-M^{2}\right)\left(p-p_{\infty}\right)+\left(1-M\right)\mathcal{T}_{4}+\mathcal{V}_{4},\label{eq:outflowL}\end{equation}
where $M$ is a reference Mach number. The terms $\mathcal{T}_{4}$ and $\mathcal{V}_{4}$ contain respectively the transverse convective and viscous terms associated with the incoming acoustic wave, and are given by
\begin{subequations}\begin{align}\mathcal{T}_{4}&=-\frac{1}{2}\left({u}_{\eta}\frac{\partial{p}}{\partial{\eta}}+p\frac{\partial{{u}_{\eta}}}{\partial{\eta}}+\rho{c}{u}_{\eta}\frac{\partial{u}_{\xi}}{\partial{\eta}}\right)\\\mathcal{V}_{4}&=\frac{\mu{c}}{2}\left[\frac{4}{3}\frac{\partial^{2}{u}_{\xi}}{\partial{\xi}^{2}}+\frac{\partial^{2}{u}_{\xi}}{\partial{\eta}^{2}}+\frac{1}{3}\frac{\partial^{2}{u}_{\eta}}{\partial{\xi}\partial{\eta}}\right].\end{align}\end{subequations}
The first term in~\eqref{eq:outflowL} is a relaxation term, which introduces partial acoustic reflections, in return for ensuring the pressure on the boundary tracks a desired far field pressure $p_{\infty}$. If this term is set to zero, the perfectly reflecting boundary condition is recovered, whilst in the limit of this term being large, $p=p_{\infty}$ on the boundary. We set this term with the value $0.278$, which was determined as an optimal coefficient in~\cite{yoo_2005}. The term $\left(1-M\right)$ in~\eqref{eq:outflowL} is a similar relaxation term applied to the transverse convective terms, and ensures that the outflow tracks a zero-vorticity condition on average. Additionally we specify the viscous condition $\partial{\tau}_{\eta\xi}/\partial{\eta}=0$ on the outflow boundary.

On inflow boundaries $\mathcal{L}_{1}$ is outgoing, whilst $\mathcal{L}_{2}$ to $\mathcal{L}_{4}$ are incoming. Following~\citet{yoo_2005} we set $\mathcal{L}_{1}$ according to its definition in~\eqref{eq:Ldef}, and specify
\begin{subequations}\begin{align}
\mathcal{L}_{3}&=\left(u_{\xi}-u_{\xi,IN}\right)\frac{0.278c}{L}+\mathcal{T}_{3}+\mathcal{V}_{3}\\
\mathcal{L}_{4}&=\left(u_{\eta}-u_{\eta,IN}\right)\frac{0.278\rho_{0}c^{2}}{L}\left(1-M^{2}\right)+\mathcal{T}_{4}+\mathcal{V}_{4}\end{align}\end{subequations}
where $\rho_{0}$ is a reference density, $u_{\eta,IN}$ and $u_{\xi,IN}$ are the desired inflow velocity components. The transverse and viscoust terms $\mathcal{T}_{3}$ and $\mathcal{V}_{3}$ are given by
\begin{subequations}\begin{align}\mathcal{T}_{3}&=-\left(u_{\eta}\frac{\partial{u}_{\eta}}{\partial\eta}+\frac{1}{\rho}\frac{\partial{p}}{\partial\eta}\right)\\
\mathcal{V}_{3}&=\frac{\mu}{\rho}\left[\frac{4}{3}\frac{\partial^{2}{u}_{\eta}}{\partial{\eta}^{2}}+\frac{\partial^{2}{{u}_{\eta}}}{\partial{\xi}^{2}}+\frac{1}{3}\frac{\partial^{2}{u}_{\xi}}{\partial{\xi}\partial{\eta}}\right]\end{align}\end{subequations}
In addition, we specify the viscous condition $\partial{\tau}_{\eta\eta}/\partial{\eta}=0$ in the inflow boundary.

\section{Numerical results for isothermal flows}\label{numres}

In this section we demonstrate our method on a range of flows. First, we validate the implementation against the simple canonical flows of Taylor-Green vortices, Poiseuille flow, and vortex convection. Subsequently we demonstrate the ability of the method to accurately model more complex flows with a study of flows past a cylinder. Finally, to show the potential of LABFM to simulate truly complex geometries, we present simulations of flow through a porous media.

\subsection{Taylor-Green vortices}

We first test the method on a simple flow - Taylor Green vortices - on a doubly periodic domain. The lack of boundaries allows an assessment of the spatial discretisation scheme in isolation. Furthermore, in the limiting case of infinite $c^{2}$ (incompressibility), a well known analytical solution exists. The computational domain is a square with periodic boundaries and side length $H$. The domain is descretised with a uniform resolution $s_{i}=s$, as described in Section~\ref{varres}. The initial conditions and solution (in the incompressible limit) are given by
\begin{subequations}\begin{align}u=&-e^{bt}\cos\left(2\pi{x}/H\right)\sin\left(2\pi{y}/H\right)\\
v=&e^{bt}\sin\left(2\pi{x}/H\right)\cos\left(2\pi{y}/H\right)\\
p=&\frac{-H}{4}e^{2bt}\left(\cos\left(4\pi{x}/H\right)+\cos\left(4\pi{y}/H\right)\right)\end{align}\end{subequations}
where the exponent $b=-8\pi^{2}H/2{Re}$, the Reynolds number is based on the undisturbed density, the initial velocity magnitude (unity), and the domain half-width $H/2$. We set $H=1$ and $Re=100$. The dimensionless time is $t^{\star}=2t/H$.
As a measure of the accuracy of our numerical solution, we take the $L_{2}$ norm of the relative error in velocity magnitude, at $t^{\star}=2$. The isentropic compressibility of the flow is given by $\beta=1/\rho_{0}{c}^{2}$, and in the present case, where we set the reference density to $\rho_{0}=1$, the compressibility is related to the maximum Mach number during the simulation by $\beta=Ma^{2}$. Although the analytical solution assumes incompressible flow, the discrepency with the compressible flow solution reduces with reducing $\beta$, allowing us to demonstrate the accuracy of our spatial discretisation.

\begin{figure}
\includegraphics[width=0.49\textwidth]{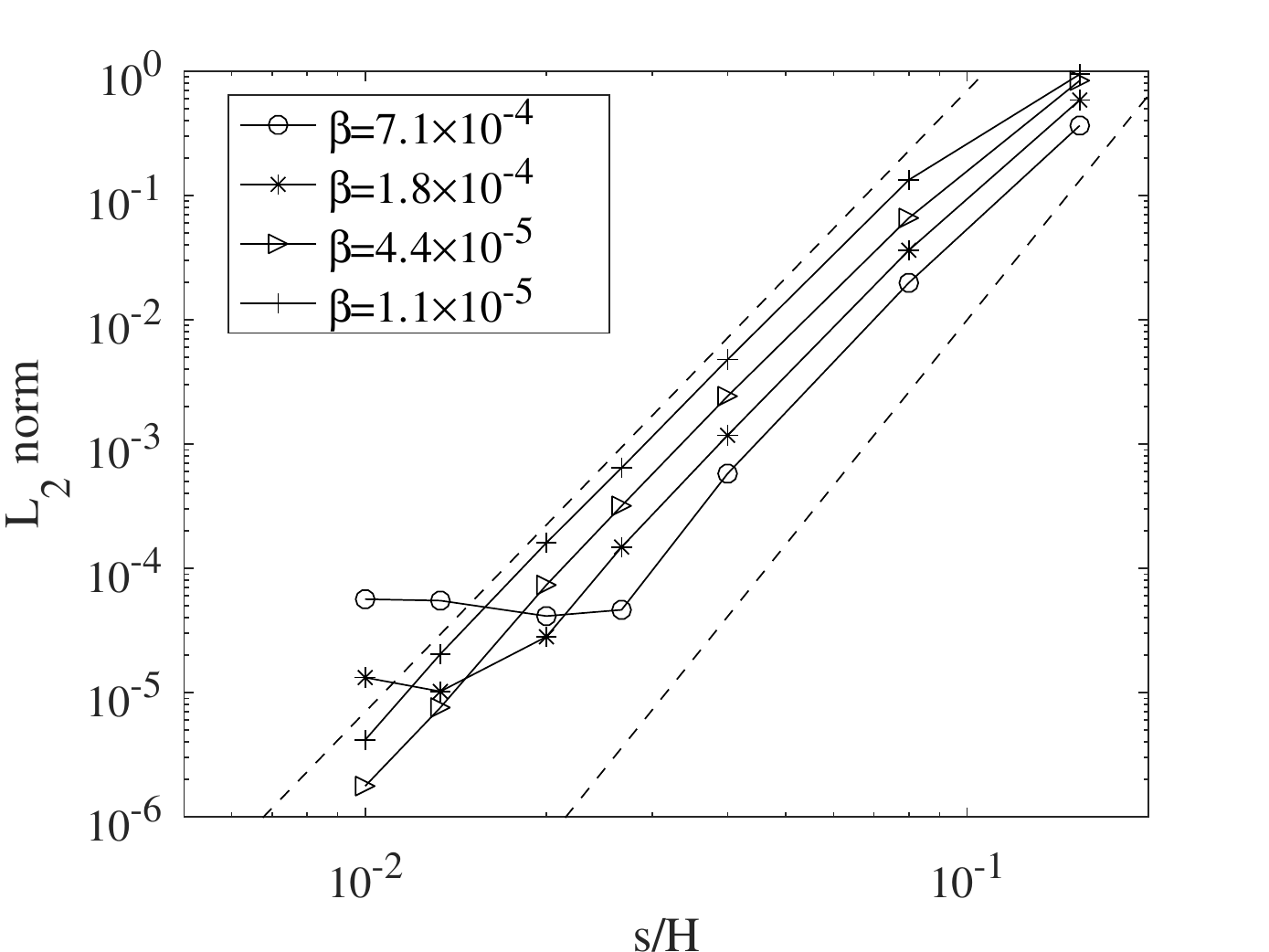}
\includegraphics[width=0.49\textwidth]{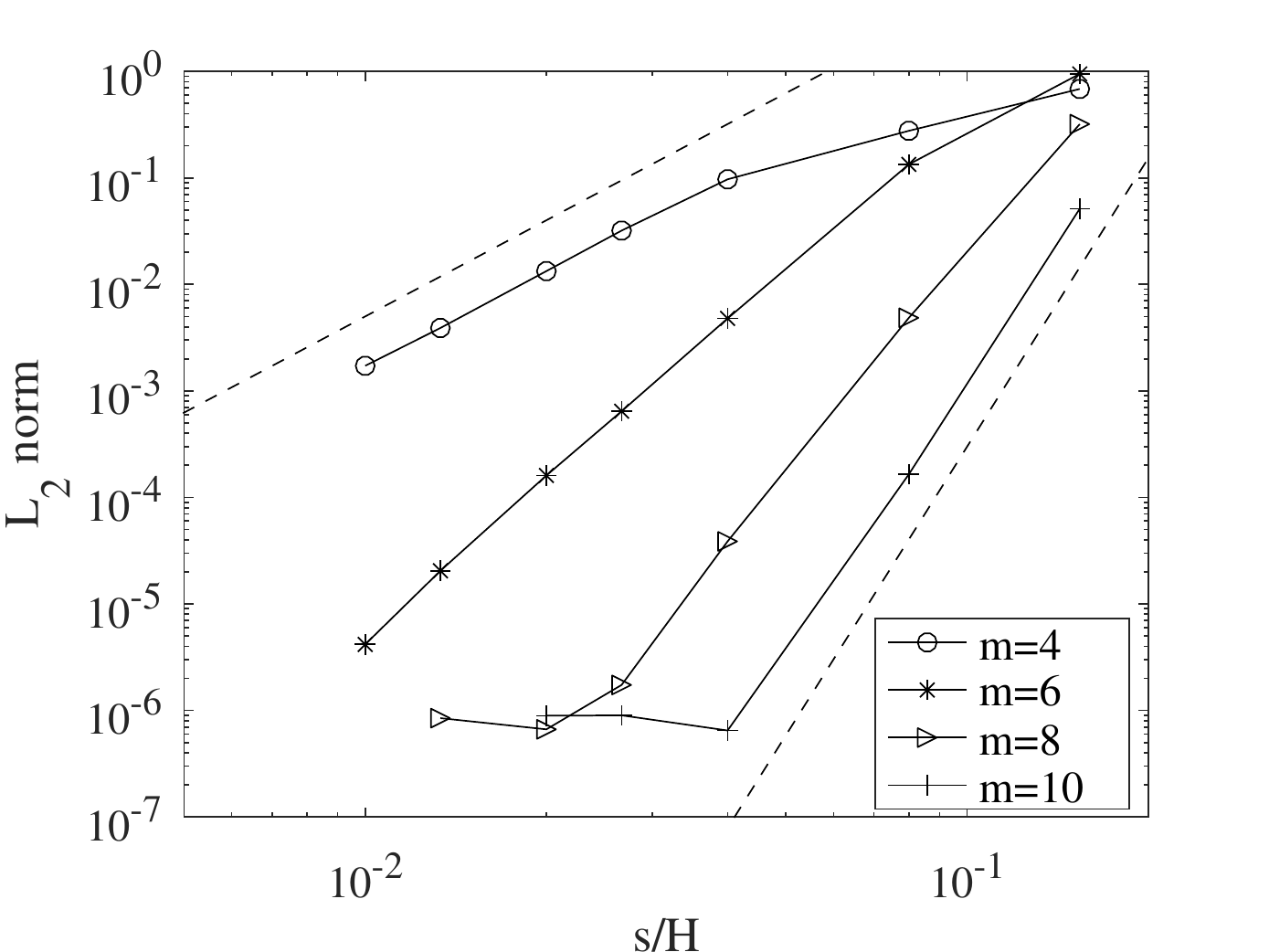}
\caption{Variation of the $L_{2}$ norm of the error in velocity at $t^{\star}=2$ for Taylor Green flow with resolution $s/H$. Left panel: various values of compressibility $\beta$, with $m=6$. The upper and lower dashed lines correspond to convergence rates of $5$ and $6$ respectively. Right panel: several values $m$, with $\beta=1.1\times10^{-5}$. The dashed lines indicate convergence rates of $3$ and $9$. In all cases the Reynolds number is $Re=100$.\label{fig:TGerror}}
\end{figure}

The left panel of Figure~\ref{fig:TGerror} shows the variation of the $L_{2}$ norm of the error of our numerical solution at $t^{\star}=2$ with resolution $s/H$, for several values compressibility $\beta$, with $m=6$. In all cases, at coarse resolutions the simulation converges towards the incompressible analytic solution at a rate proportional to $s^{5}$. This convergence rate corresponds to the order of the second derivative approximation in LABFM with $m=6$. For larger $\beta$, we reach a limiting error due to the assumption of incompressibility in the analytic solution. As $\beta$ is reduced, this error limit reduces. The magnitude of the error for a given resolution (in the convergent resolution range) reduces with increasing $\beta$. This is due to the time-step constaint, which results in $\delta{t}\propto\sqrt{\beta}$. The accumulation of errors due to time-stepping results in error magnitudes proportional to $1/\sqrt{\beta}$. The right panel of Figure~\ref{fig:TGerror} shows the variation of the $L_{2}$ norm of the error at $t^{\star}=2$ with resolution $s/H$, for a fixed value of $\beta=1.1\times10^{-5}$, as $m$ is increased. For each value of $m$, we see convergence orders of $m-1$. For $m=4$, the convergence rate is lower at coarse resolutions: this is because the effect of the filter is more significant at low resolutions and with lower $m$ - the wavelengths eliminated by the filter are closer to the wavelengths of the physical solution. For $m=8$ and $m=10$ we see convergence rates of $7$ and $9$ respectively, until a limiting error is reached, due to the assumption of incompressibility in the analytic solution.

\subsection{Poiseuille flow}

\begin{figure}
\includegraphics[width=0.6\textwidth]{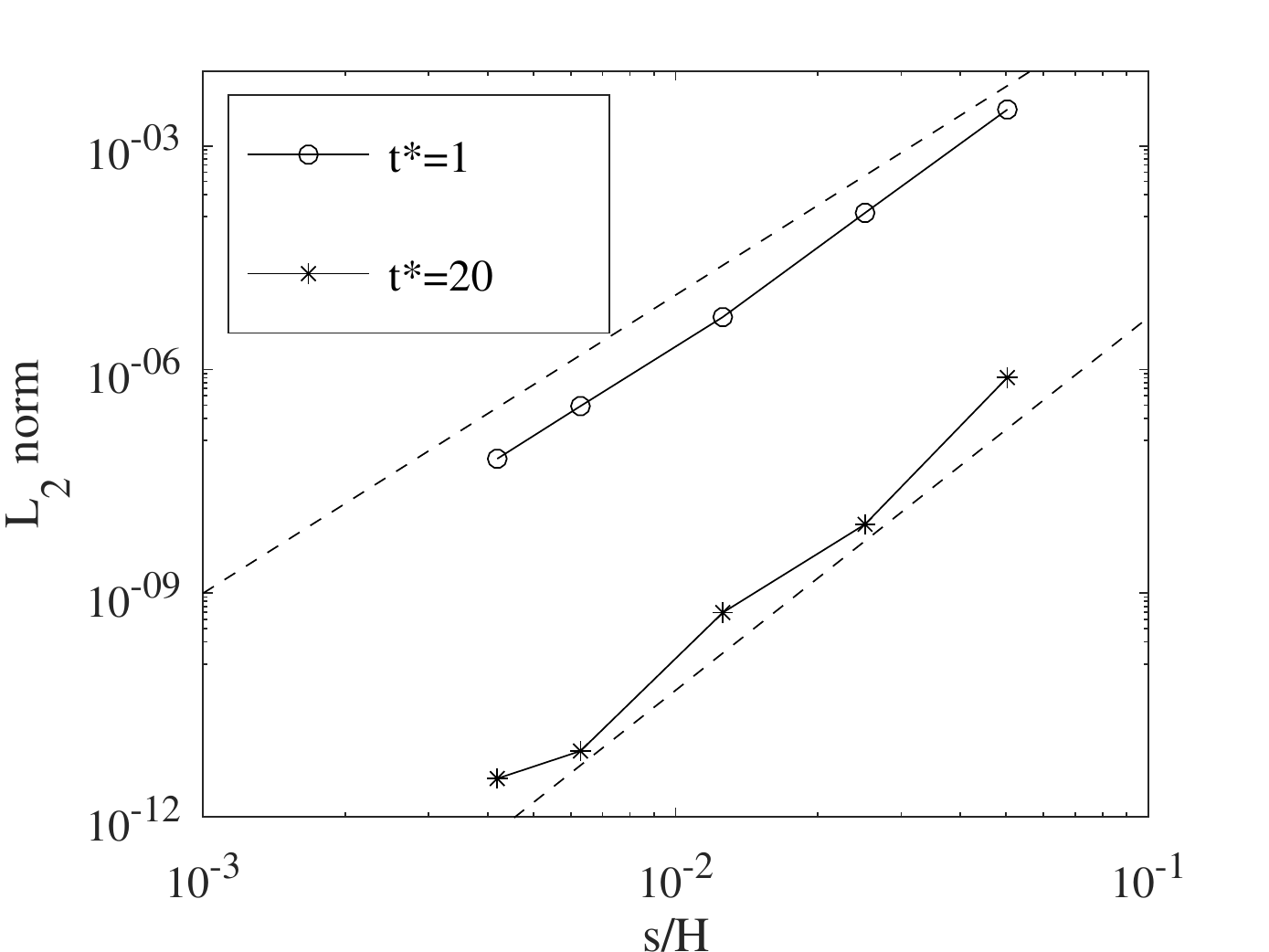}
\caption{Variation of relative $L_{2}$ norm of the error in velocity field with resolution $s/H$, for Poiseuille flow at $t^{\star}=1$ and $t^{\star}=20$. The upper and lower dashed lines correspond to convergence rates of $4$ and $5$ respectively.\label{fig:poiseuille}}
\end{figure}

This test provides a partial validation of the no slip wall boundary conditions. The domain is a square with side length $H$, and is laterally periodic, with no-slip wall conditions on the upper and lower boundaries. The flow is driven by a body force $\bm{g}=\left(g_{x},0\right)^{T}$. We non-dimensionalise with the channel width $H$ and the steady state centre-line velocity $U_{0}=\rho_{0}g_{x}H^{2}/8\mu$, with the Reynolds number $Re=HU_{0}\rho_{0}/\mu=\rho_{0}^{2}g_{x}H^{3}/8\mu^{2}=10$. The Mach number is $Ma=U_{0}/c=0.05$. We start the simulation with the fluid at rest. An anlytical solution exists for the incompressible transient start-up case (e.g. the limiting Newtonian case of the solution given in~\cite{waters_1970}), which also satisfies the isothermally compressible Navier Stokes equations; the pressure is uniform in the analytical solution. We discretise the domain with a uniform resolution $s$.  Figure~\ref{fig:poiseuille} shows the variation of the $L_{2}$ error in the velocity field with resolution at $t^{\star}=tU_{0}/H=1$ during the transient phase (circles) and at $t^{\star}=20$ once a steady state has been reached (stars). At $t^{\star}=1$ we see convergence of approximately $4^{th}$ order, whilst the steady state velocity field converges with resolution at $5^{th}$ order. Although this test case contains no acoustic energy, these results validate the ability of the wall boundary conditions to handle vorticity. 

\subsection{Lamb-Oseen vortex propagation}\label{oseen}

\begin{figure}
\includegraphics[width=0.99\textwidth]{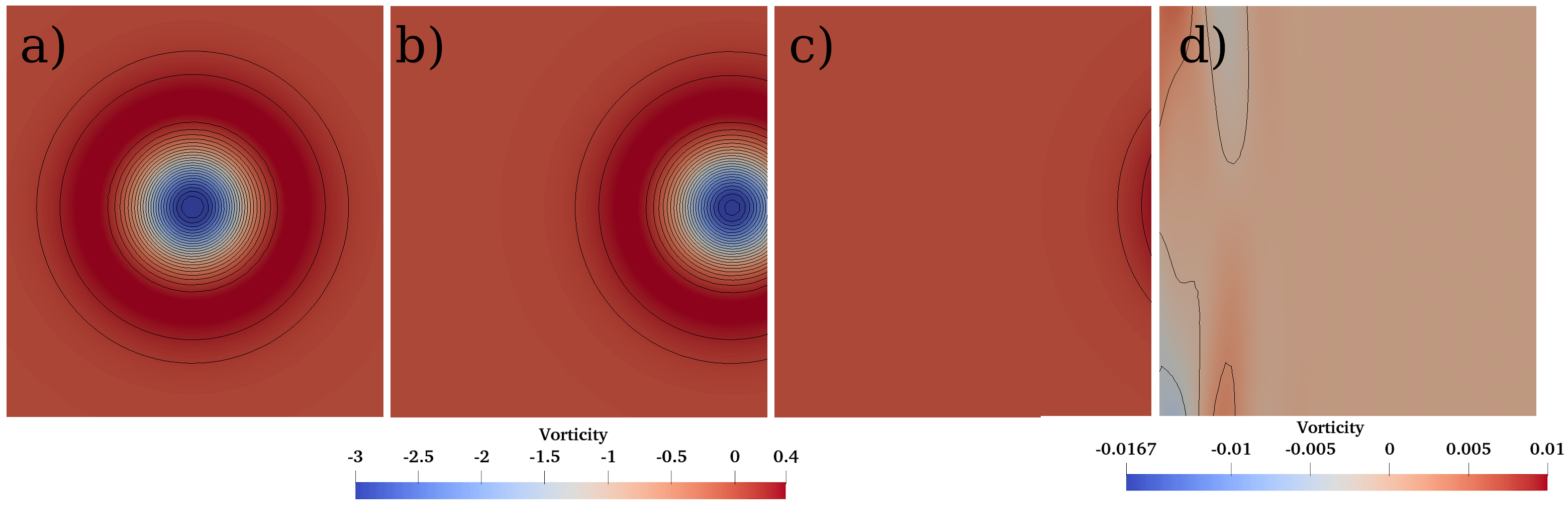}
\caption{Vorticity field at the outflow as a Lamb-Oseen vortex propagates out of the domain. The region shown is $\left(x,y\right)\in\left(\left[H/8,H/2\right],\left[-3H/16,3H/16\right]\right)$. Frames a) to d) correspond to times $t^{\star}=6$, $t^{\star}=9$, $t^{\star}=12$, and $t^{\star}=15$ respectively. In frame d), the vortex has left the domain, and a different colour and contour scale is used to show the residual vorticity field due to the vortex passing through the boundary.\label{fig:lo}}
\end{figure}

Here we briefly demonstrate the transparency of the non-reflecting outflow boundary conditions to both acoustic waves and vorticity. The test consists of a Lamb-Oseen vortex superimposed on a uniform flow. The domain is square with size length $H$, given by $\left(x,y\right)\in\left(\left[-H/2,H/2\right],\left[-H/2,H/2\right]\right)$. The discretisation has a uniform resolution, with nodal disorder created as described in Section~\ref{varres}. At the left and right boundaries we apply the partially reflecting inflow and outflow conditions respectively. The upper and lower boundaries are periodic. The flow is uniform with velocity $U_{0}$, and a vortex positioned at $x=y=0$ at time $t=0$. The initial conditions are
\begin{equation}\left(u,v\right)^{T}=\left(U_{0},0\right)^{T}+\frac{1}{\rho_{0}}\left(\frac{Cy}{R_{c}^{2}}\exp\left(-\frac{\left(x^{2}+y^{2}\right)}{2R_{c}^{2}}\right),-\frac{Cx}{R_{c}^{2}}\exp\left(-\frac{\left(x^{2}+y^{2}\right)}{2R_{c}^{2}}\right)\right)^{T},\end{equation}
where $R_{c}$ is the radius of the vortex and $C$ is the vortex strength. The initial density field is given by
\begin{equation}\rho=\rho_{0}\left\{1-\frac{C^{2}}{c^{2}R_{c}^{2}}\exp\left(-\frac{\left(x^{2}+y^{2}\right)}{2R_{c}^{2}}\right)\right\}.\label{eq:oseenrho}\end{equation}
We set the free stream Mach number $Ma=U_{0}/c=0.05$, the vortex radius $R_{c}=H/25$, and the vortex strength $C=cH/300$. The vortex Reynolds number is $Re=U_{0}R_{c}\rho_{0}/\mu=7500$. This (dimensionless) vortex strength results in a minimum streamwise velocity of $0$ at time $t=0$. By the time the vortex has reached the outflow boundary, its strength has reduced due to viscous dissipation, but this still presents a challenging test of the boundary conditions. Although~\eqref{eq:oseenrho} corresponds to the initial pressure field used by other authors~\cite{poinsot_1992,yoo_2007}, those other works considered the fully compressible Navier Stokes equations, and for this test case imposed uniform initial density. We are considering isothermal flows, and this constraint results in the initial density field in~\eqref{eq:oseenrho} being out of equilibrium with the velocity field. As a result, an acoustic wave is generated at $t=0$ which propagates out through the non-reflecting boundaries. By the time the vortex reaches the outflow boundary, the acoustic energy has largely left the domain, and the density field in the vortex is steady, demonstrating the transparency of the boundary conditions to acoustic waves. Figure~\ref{fig:lo} shows the vorticity field as the vortex propagates out of the domain, with a resolution of $R_{c}/s=12$. Frame a) corresponds to dimensionless time $t^{\star}=tU_{0}/R_{c}=6$, just as the vortex approaches the outflow boundary, and frames b) to d) follow at $t^{\star}=9$, $t^{\star}=12$, and $t^{\star}=15$. In frames b) and c), as the vortex is passing through the boundary, we can clearly see the contours of vorticity remain circular. This absence of distortion demonstrates the transparency of the boundary to vorticity. In frame d) of Figure~\ref{fig:lo}, the vortex has left the domain, and we use a different colour and contour scale to highlight the residual vorticity in the domain. The magnitude of the residual vorticity is small, below $1\%$ of the strength of the vortex, again indicating that the passage of the vortex through the artificial boundary generates only minimal vorticity within the domain.

\begin{figure}
\includegraphics[width=0.49\textwidth]{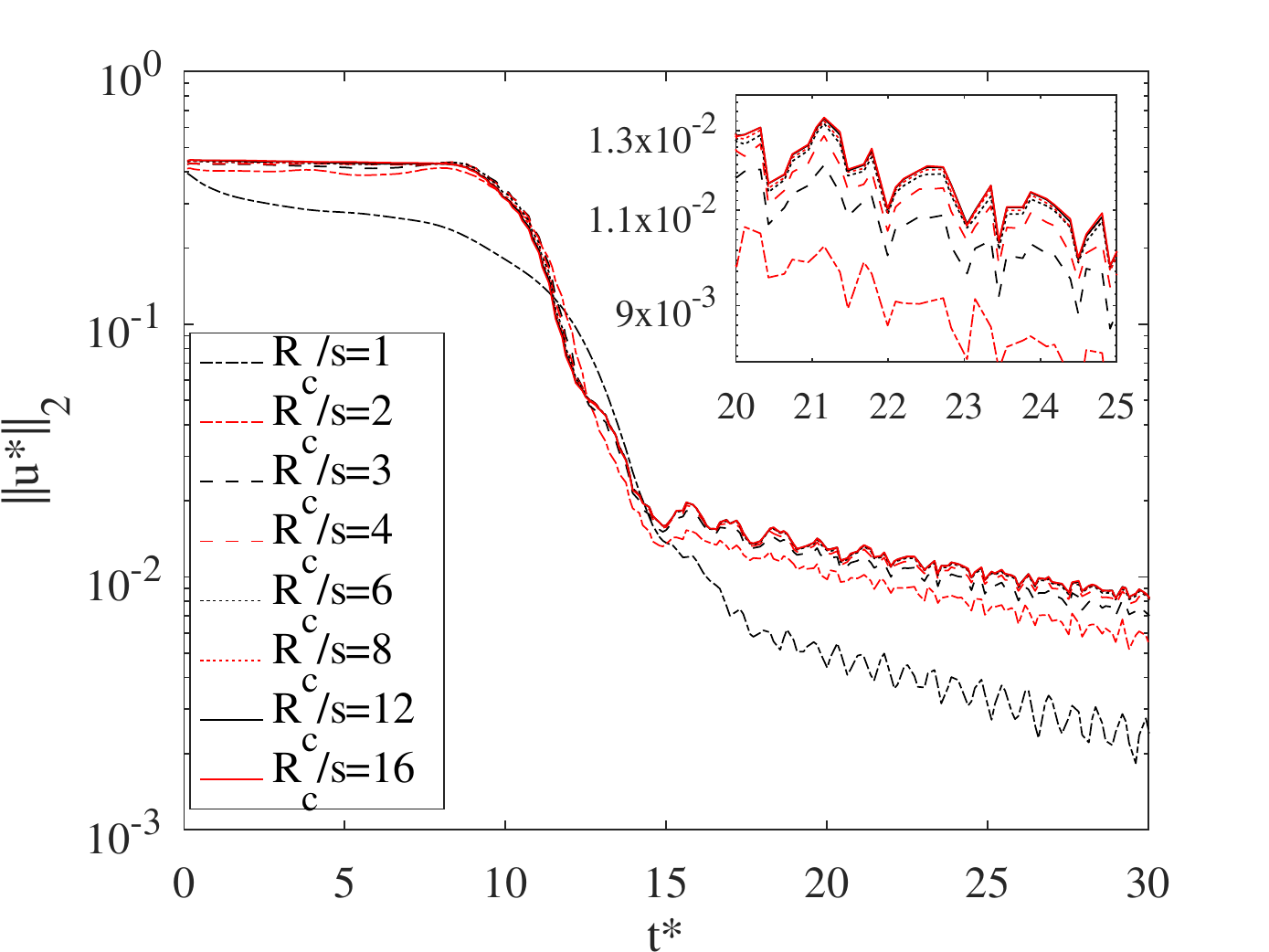}
\caption{Variation of the $L_{2}$ norm of the velocity disturbance $\left(u-U_{0},v\right)^{T}/U_{0}$ with dimensionless time $t^{\star}=tU_{0}/R_{c}$, for different resolutions, for the vortex advection test problem.\label{fig:oseen_vel}}
\end{figure}

We now denote the dimensionless disturbed velocity as $\bm{u}^{\star}=\left(u-U_{0},v\right)^{T}/U_{0}$, such that in the limit of large $t$, $\bm{u}^{\star}=\bm{0}$. Figure~\ref{fig:oseen_vel} shows the temporal variation of the $L_{2}$ norm of $\bm{u}^{\star}$ for several resolutions (denoted in the legend by $R_{c}/s$, the number of nodes across the vortex radius). For all resolutions, we see a drop in the velocity as the vortex leaves the domain between $t^{\star}=9$ and $t^{\star}=15$, followed by a gradual decay of the velocity field towards $\bm{u}^{\star}=\bm{0}$. For the coarse resolutions, the vortex is under-resolved, resulting in significant deviations compared with the finer resolutions. As the resolution is refined, we see clear convergence (visible in the inset in Figure~\ref{fig:oseen_vel}) towards the finest resolution. We calculate convergence rate (with resolution) towards the results with $R_{c}/s=16$, and find that for resolutions $R_{c}/s=\left[3,4,6,8,12\right]$ the convergence rate approaches third order ($\left[1.43,2.31,2.98,3.00,2.84\right]$). This rate is consistent with the order of the second derivative approximations used on the domain boundary.

\subsection{Flow past an impulsively started cylinder}

We next consider the problem of flow past a cylinder, which has been extensively studied by various authors in the context of early stage vortex development(e.g.~\cite{bouard_1980,smith_1988,li_2004}), and long-term behaviour (e.g.~\cite{don_1990,burbeau_2002,taira_2007,liu_2010,canuto_2015}). The problem consists of a rectangular domain $\left(x,y\right)=\left[-\Lambda_{u}D,\Lambda_{d}D\right]\times\left[-\Lambda_{w}D/2,\Lambda_{w}D/2\right]$, with a solid cylinder of diameter $D=0.25$ placed at the origin. In all cases, we set the upstream reach to $\Lambda_{u}=4$. We vary the downstream reach $\Lambda_{d}$ and the width $\Lambda_{w}$. We impose periodic conditions at the upper and lower boundaries. We impose inflow and outflow conditions at the left and right boundaries respectively. We initialise the simulation with a uniform density $\rho_{0}$ everywhere, and a uniform velocity $\bm{u}=\left(U_{0},0\right)^{T}$ everywhere except the cylinder surface, where the velocity is zero. We non-dimensionalise the problem with the inflow velocity $U_{0}$ and the cylinder diameter, yielding a Reynolds number $Re=\rho_{0}U_{0}D/\mu$, and a dimensionless time of $t^{\star}=U_{0}t/D$. These initial conditions correspond to an impulsively started cylinder, and in this section we will consider both the early stage wake development and the long term steady or periodic flow behaviour.

\subsubsection{Steady flow at $Re=50$}

\begin{figure}
\includegraphics[width=0.99\textwidth]{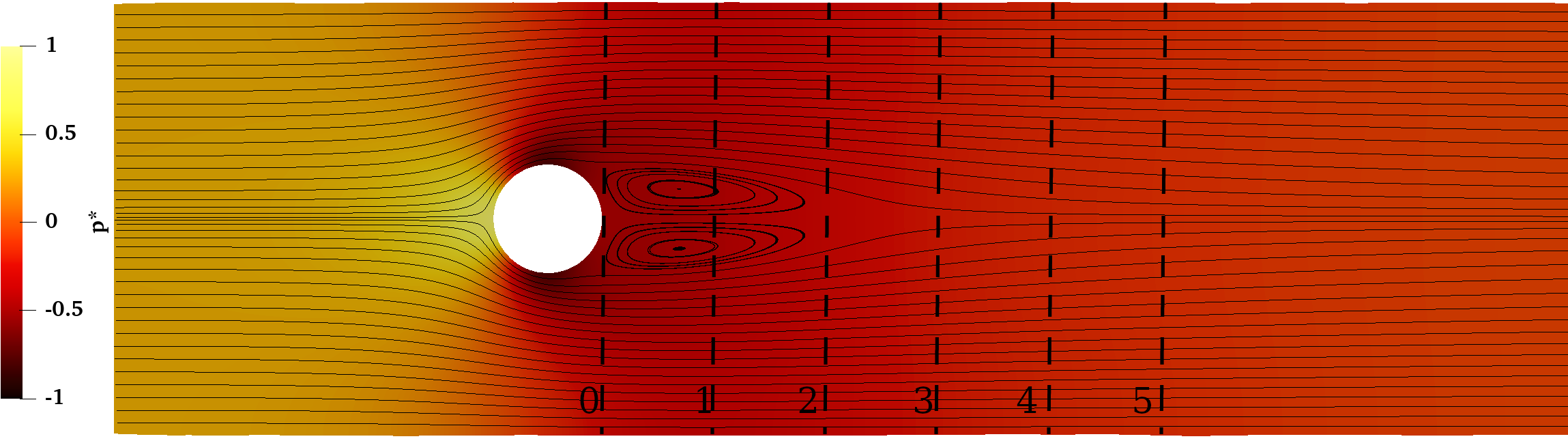}
\caption{Streamlines and dimensionless pressure field $p^{\star}=\left(p-c^{2}\rho_{0}\right)/\rho_{0}U_{0}^{2}$ for the steady state flow around a cylinder with $Ma=0.05$ and $Re=50$. The vertical dashed lines indicate the locations at which streamwise velocity profiles are calculated for comparison with the results of~\citet{zhang_2006}. Note, we do not show the entire domain, only the region $x<9.4D$.\label{fig:cyl_ss}}
\end{figure}

\begin{figure}
\includegraphics[width=0.6\textwidth]{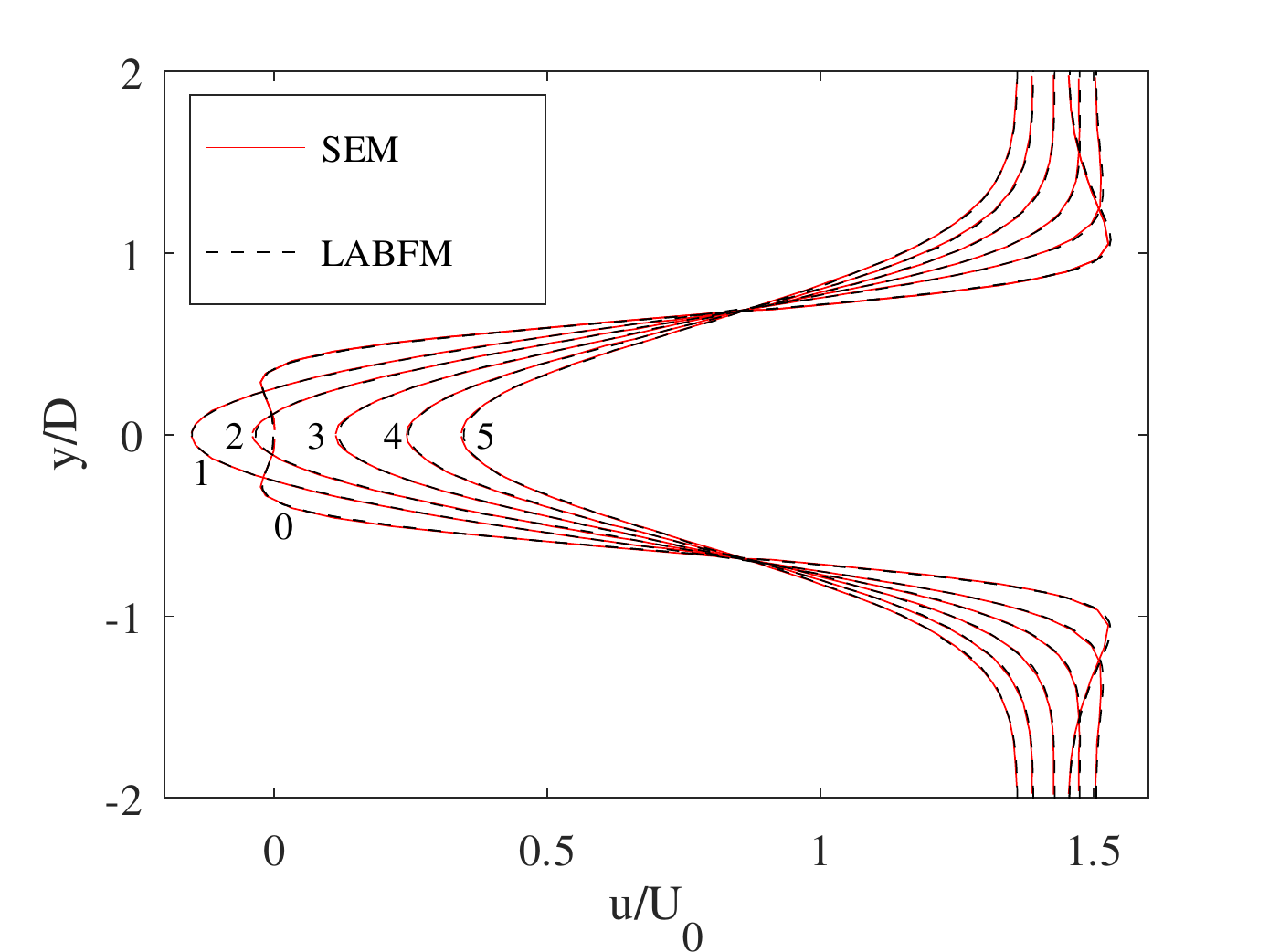}
\caption{Streamwise velocity profiles at the downstream edge of the cylinder ($x=D/2$) and at $5$ locations downstream ($x=D/2+\left[D,2D,3D,4D,5D\right]$), obtained with LABFM (dashed black lines), compared with the SEM results of~\citet{zhang_2006} (red lines). The numerical annotations indicate the distance (in cylinder diameters) downstream of the trailing edge for each profile.\label{fig:cyl_prof}}
\end{figure}

%% Zhang 2006, Spectral element comparisons...
We first provide further validation of the accuracy of our approach by comparing with published numerical results for a configuration with a stable long term solution. We set $\Lambda_{d}=28$, $\Lambda_{w}=4$, the Reynolds number $Re=50$ and the Mach number $Ma=U_{0}/c=0.05$. For this Mach number, the maximum density variation in the flow is less than $0.1\%$, allowing for comparisons with the numerical results of~\cite{zhang_2006}, who used a spectral element method (SEM) to solve the incompressible Navier Stokes equations. For this comparison we set the resolution at the cylinder wall to $s_{min}=D/50$, and the inflow/outflow resolution to $s_{max}=D/25$. Within a distance $D$ from the cylinder wall the resolution is uniform, after which it increases linearly to $s_{max}$ at a distance $3D$ from the cylinder. The resolution is again uniform at distances greater than $3D$ from the cylinder. With this node distribution, we obtain a converged solution, to within $0.03\%$. We run the simulation until $t^{\star}=80$, by which time a steady state flow field is obtained, shown in Figure~\ref{fig:cyl_ss}. We calculate the streamwise velocity profiles at the downstream edge of the cylinder ($x=D/2$), and at $5$ locations downstream of the cylinder, with coordinates $x=D/2+\left[D,2D,3D,4D,5D\right]$, indicated by the vertical dashed lines labelled $0$ to $5$ on Figure~\ref{fig:cyl_ss}. Figure~\ref{fig:cyl_prof} shows the velocity profiles at these locations, with our results (dashed black lines) compared with the numerical results of~\cite{zhang_2006} (solid red lines). In all profiles, we see a very close match between our results and the SEM results of~\cite{zhang_2006}. This provides further validation of the accuracy of our method, and of the implementation of the boundary conditions framework described in~\ref{bound}. 

\begin{figure}
\includegraphics[width=0.9\textwidth]{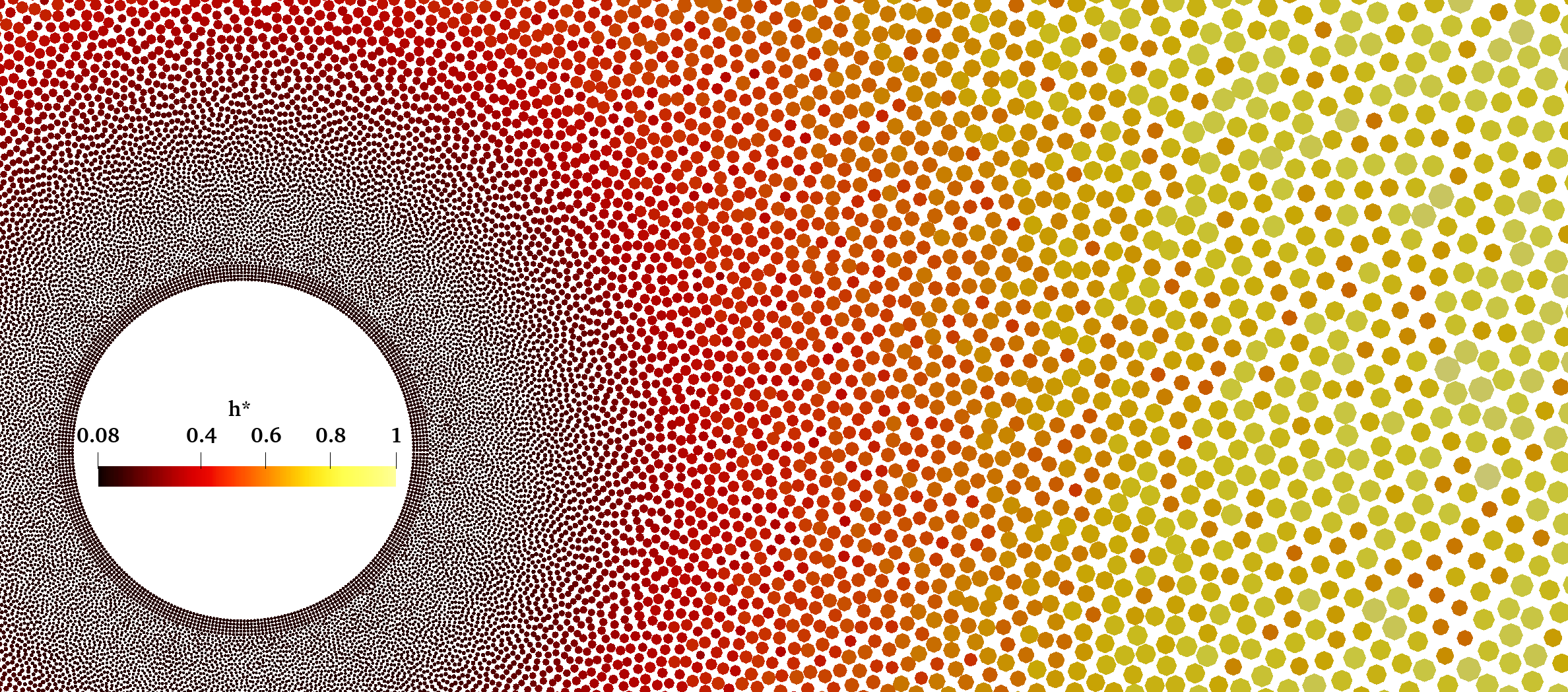}
\caption{Node distribution around the cylinder. The colour and size of nodes indicate the non-dimensional stencil lengthscale $h^{\star}=h/h_{max}$. Note the stencil optimisation procedure results in a non-smooth variation of $h^{\star}$.\label{fig:cyl_nodes}}
\end{figure}

\subsubsection{Early stage wake development at $Re=550$}

\begin{figure}
\includegraphics[width=0.7\textwidth]{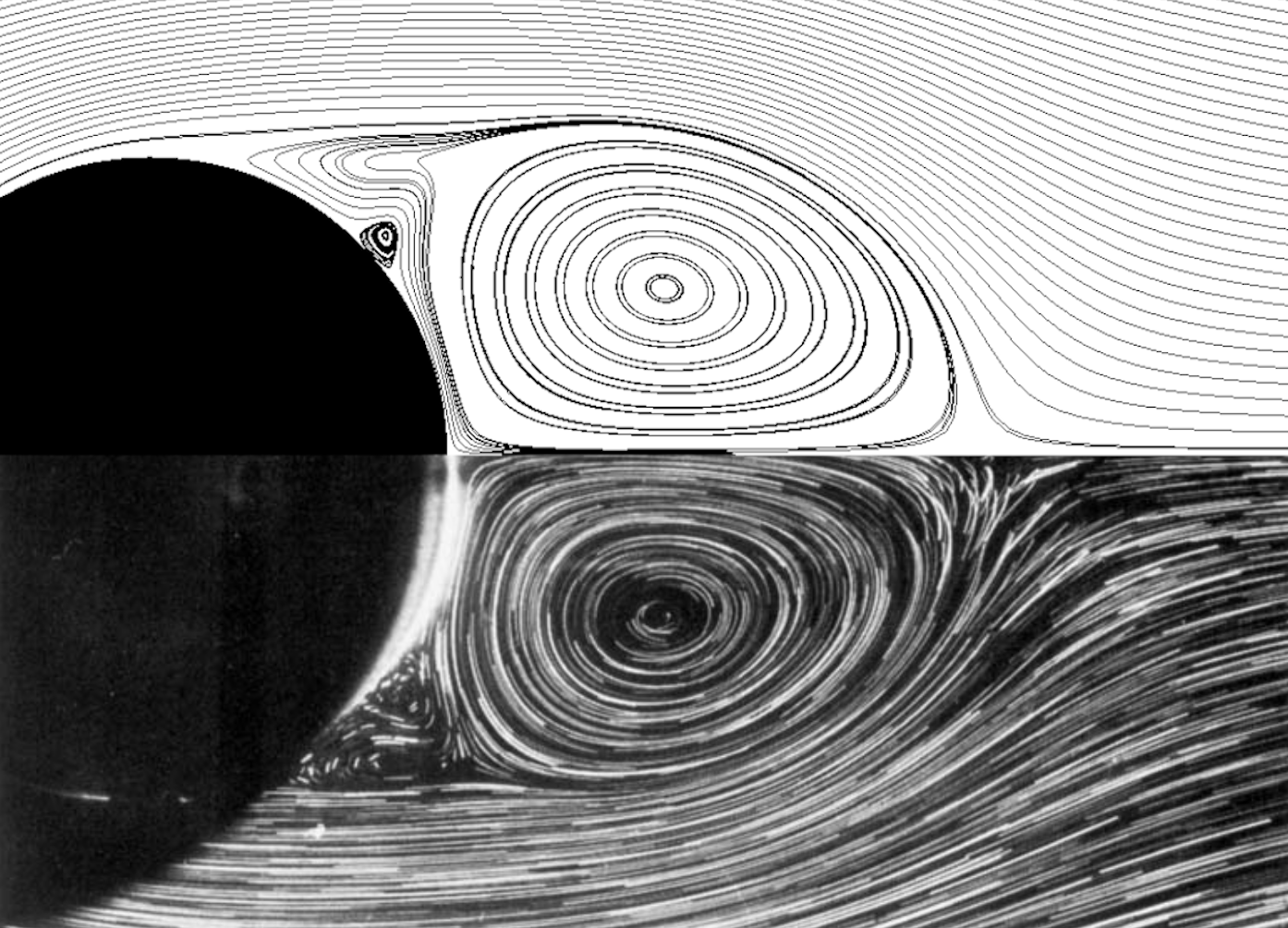}
\caption{Flow past an impulsively started cylinder at $Re=550$ and $t^{\star}=2.5$. The upper half of the image shows streamlines obtained from our numerical simulation. The lower half shows streaklines obtained by experiment~\cite{bouard_1980}\label{fig:re550_cyl}}
\end{figure}

%% Early stage
We next look at the early stage development of the vortical structures in the wake. For this case, we use a low Mach number of $Ma=1/75$, and set the Reynolds number to $Re=550$, to match the experimental work of~\cite{bouard_1980}. We increase the width of the domain, setting $\Lambda_{w}=10$, such that the effects of the upper and lower boundaries are negligible. To resolve the thin boundary layer at this higher Reynolds number, we further refine the node distribution around the cylinder. The finest resolution $s_{min}$ occurs at the cylinder wall, with the resolution at the inflow and outflow $s_{max}=12{s}_{min}$. The resolution is varied smoothly with distance from the cylinder surface. Between $0.2D$ and $D$ from the cylinder a cosine function is used to transition from $s_{min}$ to $4s_{min}$, beyond which $s$ increases linearly to $s_{max}$ at $3D$ from the cylinder. Figure~\ref{fig:cyl_nodes} shows the node distribution, with nodes coloured and scaled by the stencil-scale $h$ after the stencil optimisation procedure has been applied. Figure~\ref{fig:re550_cyl} provides a comparison between our numerical results (streamlines, upper half), and the experimental work (streaklines, lower half) of~\cite{bouard_1980}, at time $t^{\star}=2.5$. We see a qualitatively good match with experimental results. The primary vortex length is accurately reproduced, and we capture the formation of a secondary recirculation region, and the bulge in the primary vortex. 
The ability of LABFM to incorporate spatially varying resolution is particularly valuable here. We require a fine resolution at the cylinder wall to resolve the thin vorticity sheet which is generated during the early stages (from a velocity discontinuity at $t=0$), whilst to ensure that boundary effects don't contaminate the solution, we require a reasonably large domain, much of which contains a relatively uninteresting flow (i.e. almost uniform flow far from the cylinder). We also note that the use of a compressible formulation is likely to aid the simulation of this problem, as the compressibility allows the energy associated with the initial velocity discontinuity to be dissipated over a finite time, in the form of acoustic energy.

\subsubsection{Vortex shedding at $Re=1000$ and non-neglible $Ma$}

As a final test for the case of flow past a cylinder we increase the compressibility, setting $Ma=0.2$, with $\beta=1/25$. 
In the context of compressible flows, this geometry has been investigated with a range of methods. The problem was studied in~\cite{don_1990} with a spectral method at $Ma=0.4$ and $Re=80$. In~\cite{burbeau_2002}, a discontinuous Galerkin method was used to model the flow with $Ma=0.2$ and $Re=100$. The effects of Mach number were studied by~\citet{canuto_2015} for a range of $Re\in\left[20,100\right]$ and $Ma\in\left[0,0.5\right]$. \citet{liu_2010} used a high order immersed boundary method with $Ma=0.2$ and $Re=10^{3}$ to demonstrate their artificial boundary conditions. The flow in~\cite{liu_2010} is nearly isothermal, with temperature variation less than $1\%$, and so we take the compressible case with $Ma=0.2$ as a good problem to test the qualitative behaviour of our method. 

\begin{figure}
\includegraphics[width=0.99\textwidth]{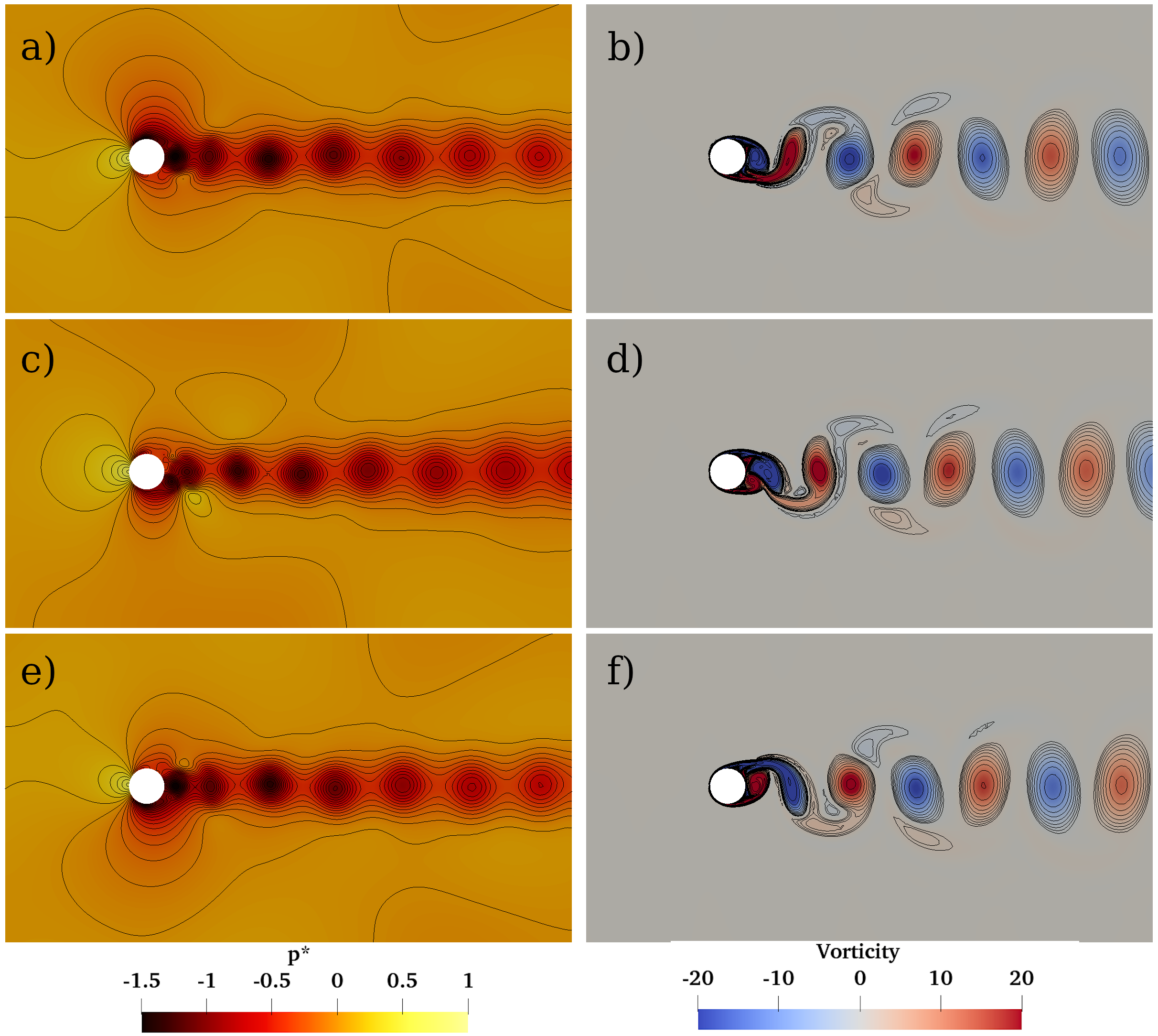}
\caption{Flow past a cylinder at $Re=550$ and $Ma=0.2$. a,c,e): Colour and contours indicate the dimensionless pressure $p^{\star}=\left(p-c^{2}\rho_{0}\right)/\rho_{0}U_{0}^{2}$. b,d,f): Colour and contours indicate the vorticity field. The frames correspond to times a,b) $t^{\star}=150$, c,d) $t^{\star}=151$, and e,f) $t^{\star}=152$. Note the vorticity contours are distributed uniformly over $\ln\left\lvert\omega\right\rvert$ for $\left\lvert\omega\right\rvert\ge1$.\label{fig:re550_cyl_long}}
\end{figure}

We increase the Reynolds numbers to $Re=10^{3}$, matching the parameters studied in~\cite{liu_2010}. Figure~\ref{fig:re550_cyl_long} shows the pressure (left: a,c,e) and vorticity (right: b,d,f) fields around the cylinder late in the simulation (from $t^{\star}=150$). By this time, the vortex street behind the cylinder is established and steady. In frames a) and b) of Figure~\ref{fig:re550_cyl_long}, a vortex is about to leave the domain, whilst in frames c) and d) the same vortex is passing through the outflow boundary, and in frames e) and f) it has left the domain. Qualitatively, Figure~\ref{fig:re550_cyl_long} shows good agreement with the pressure and vorticity profiles shown in~\cite{liu_2010} (their Figures 7 and 10). Throughout, we see the pressure tracks the desired outflow density well, with $p^{\star}$ on the outflow boundary close to zero away from the vortices, and no visible generation of acoustic energy with the passage of vortices through the boundary. Looking at the right panels in Figure~\ref{fig:re550_cyl_long}, we see the closed contours of vorticity undergo negligible distortion as they approach and propagate through the boundary. These observations support our results in Section~\ref{oseen} in demonstrating the transparency of the outflow boundary condition to vorticity.

\subsection{Flow through a porous media}

For a final test problem, we simulate flow through a porous matrix. This case is designed to demonstrate the potential of the method for high order simulations of flows through complex geometries. The domain is rectangular, defined by $\left(x,y\right)=\left(\left[-3H/2,8H\right],\left[-H/2,H/2\right]\right)$. At the left and right boundaries inflow and outflow conditions are imposed respectively, whilst the upper and lower boundaries are periodic. The porous matrix is confined to the region $\left\lvert{x}\right\rvert<H/2$, and consists of an array of irregularly shaped solid obstacles, with characteristic diameter $D=0.16H$. The shape of the obstacles are defined by a radius which varies with polar angle as
\begin{equation}r_{o}\left(\theta\right)=\frac{D}{2}\left\{1+\displaystyle\sum_{k=1}^{5}a_{k}\sin\left[k\left(\theta-\theta_{0}\right)\right]\right\},\end{equation}
where $a_{k}$ are randomly set to $-0.1$, $0$, or $0.1$, and the random $\theta_{0}\in\left[0,2\pi\right]$ describes the orientation of the obstacle. The spatially varying resolution is set as follows. For all nodes within a distance $20s_{min}$ of a solid boundary, the resolution is $s_{min}$. For nodes between $20s_{min}$ and $100s_{min}$ of a solid boundary, the resolution varies smoothly from $s_{min}$ to $4s_{min}$. For nodes between $100s_{min}$ and $300s_{min}$ of a solid boundary, the resolution varies linearly up to $s_{max}=16s_{min}$. Elsewhere, the resolution is uniform at $s_{max}$. We obtain a resolution-converged solution with $s_{min}=H/480$. We use $D$ as the characteristic length scale, and as a characteristic velocity scale we use the inflow velocity $U_{0}$. The dimensionless time is $t^{\star}=tU_{0}/D$. We investigate several values of Reynolds number: $Re_{0}=\rho_{0}U_{o}D/\mu=\left[75, 150, 225, 300, 375, 450\right]$, and set the Mach number to $Ma_{0}=U_{0}/c=0.0375$. The flow field is impulsively started (i.e., $\bm{u}\left(t=0\right)=\left(U_{0},0\right)^{T}$ everywhere except the solid boundaries), and is run until $t^{\star}=400$. By $t^{\star}=100$ the initial transients have decayed and a statistically steady flow is established.

\begin{figure}
\includegraphics[width=0.99\textwidth]{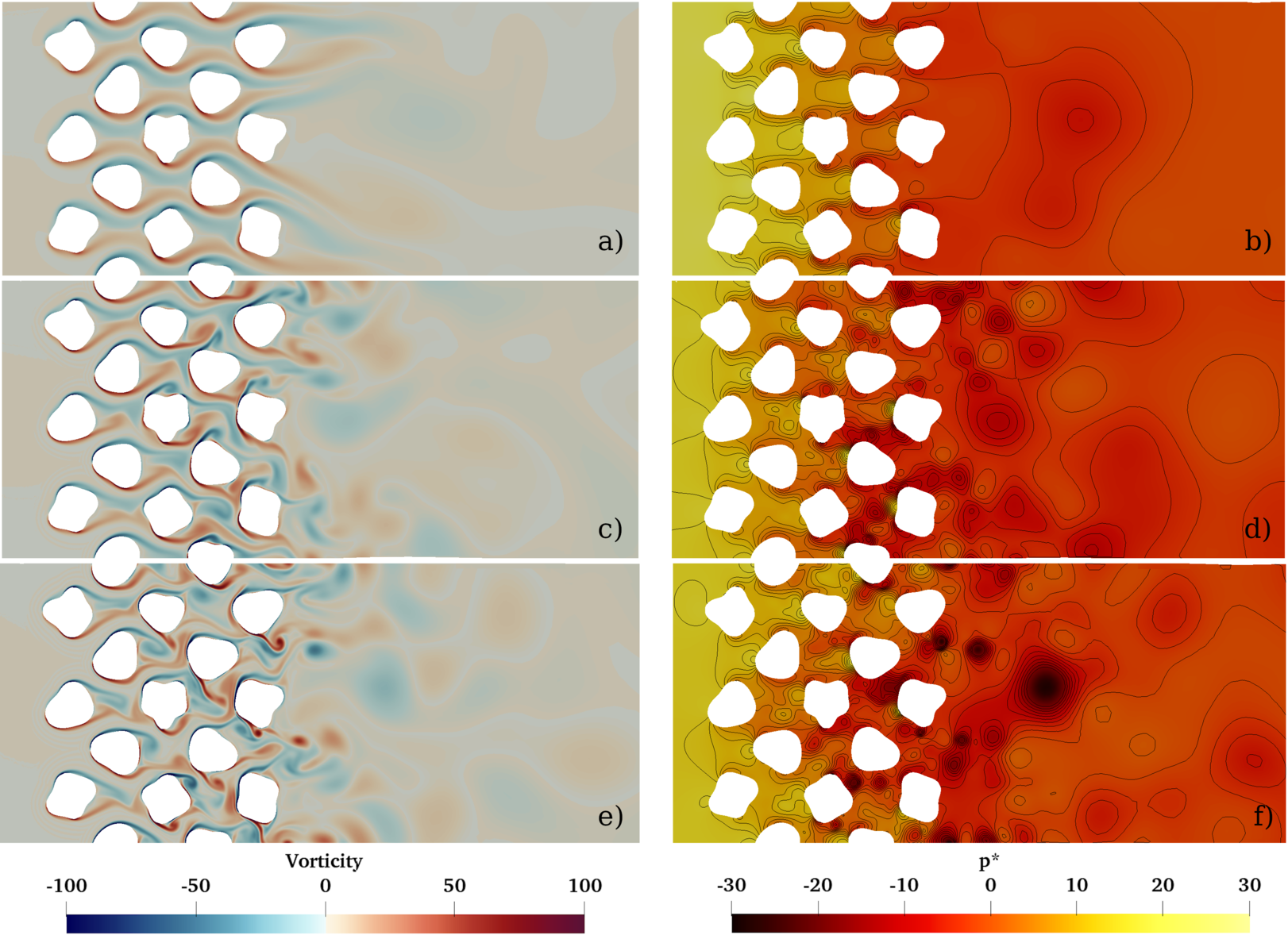}
\caption{Snapshots of the flow field through a porous band at several Reynolds numbers. a),b): $Re_{0}=75$; c),d): $Re_{0}=225$; e),f) $Re_{0}=450$. The left panels (a,c,e) show the dimensionless vorticity field. The right panels (b,d,f) show the dimensionless pressure field.\label{fig:porous1}}
\end{figure}

\begin{figure}
\includegraphics[width=0.49\textwidth]{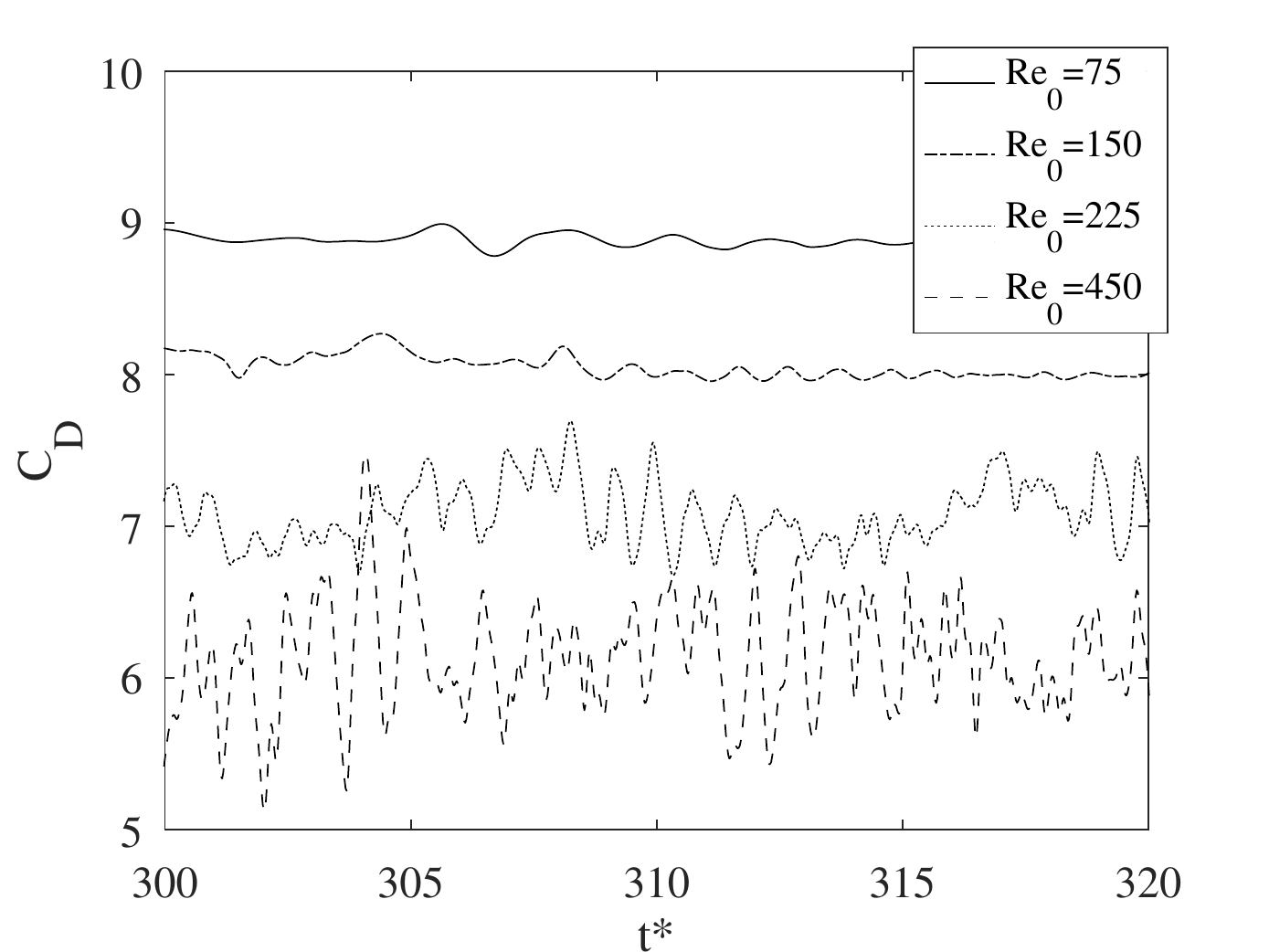}
\includegraphics[width=0.49\textwidth]{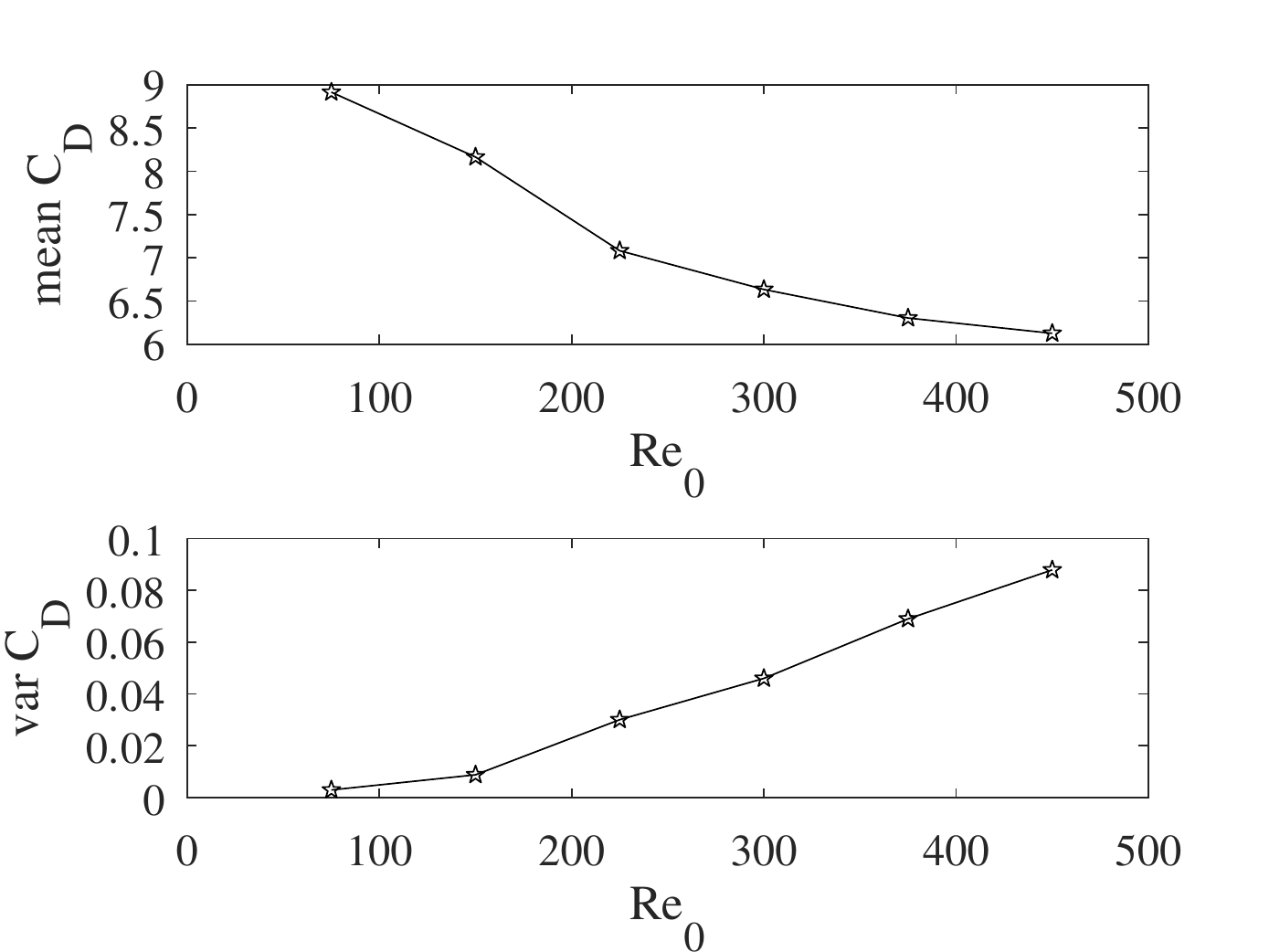}
\caption{Drag coefficients on the porous band. Left panel: Time variation of the drag coefficient for several Reynolds numbers. Right panel: mean (upper) and variance (lower) of drag coefficient for several Reynolds numbers.\label{fig:porousdrag}}
\end{figure}

We denote the intrinsic average velocity in the porous region $\left\lvert{x}\right\rvert<H/2$ as $\left\lvert\bm{U}\right\rvert$. This velocity is calculated during the statistically steady part of the simulation, and is used to obtain a pore Reynolds number $Re_{p}=\rho_{0}\left\lvert\bm{U}\right\rvert{D}/\mu$. The inflow Reynolds numbers $Re_{0}=\left[75, 150, 225, 300, 375, 450\right]$ correspond to pore Reynolds numbers of $Re_{p}=\left[142, 303, 473, 635, 815, 968\right]$ respectively. Figure~\ref{fig:porous1} shows the flow fields (after a statistically steady state has been achieved) for several Reynolds numbers. The left panels show the dimensionless vorticity $\omega{D}/U_{0}$, whilst the right panels show the dimensionless pressure field $p^{\star}=\left(p-\rho_{0}c^{2}\right)/\rho_{0}U_{0}^{2}$. We see that for $Re_{0}=75$ the unsteady vortical structures are largely confined to the region downstream of the porous band. As the Reynolds number increases, the vortices move upstream, into the pore space, increasing in intensity and reducing in size. We calculate the drag total drag on the porous band $F_{D}$, from which an average drag coefficient (per element of the porous matrix) is obtained as $C_{D}=2F_{D}/15\rho_{0}U_{0}^{2}D$. The left panel of Figure~\ref{fig:porousdrag} shows the time variation of the drag coefficient for several Reynolds numbers. With increasing Reynolds number the drag coefficient decreases, but the temporal variation increases, as the unsteady flow moves into the pore spaces. The right panel of Figure~\ref{fig:porousdrag} shows the mean and variance of $C_{D}$ over the period $t^{\star}\in\left[100,400\right]$, for a range of Reynolds numbers. The non-linear reduction in mean drag coefficient with increasing Reynolds number is clear, and in qualitative agreement with experimental results~\cite{tanino_2008} and theoretical estimates~\cite{sonnenwald_2019} for flows past random cylinder arrays, whilst above $Re_{0}=150$ we see an approximately linear increase in the variance of $C_{D}$.

\begin{figure}
\includegraphics[width=0.99\textwidth]{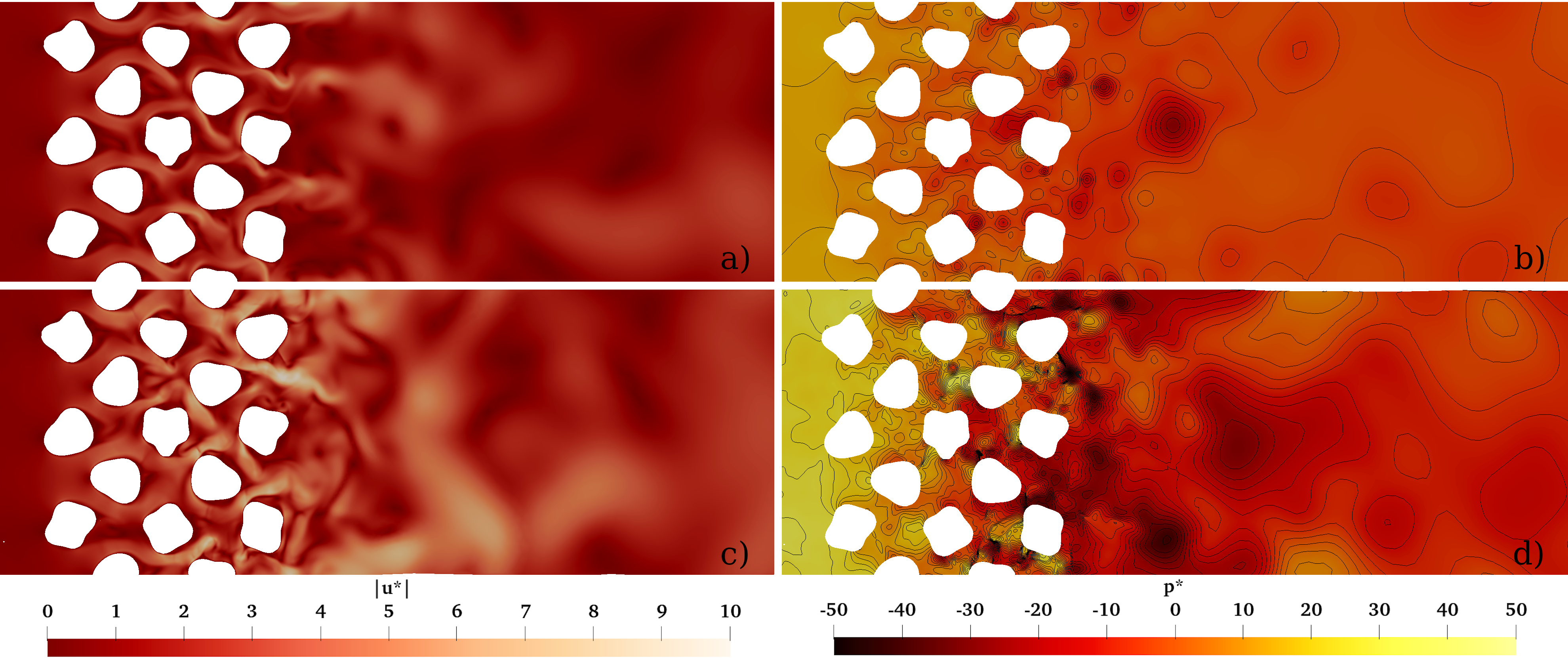}
\caption{Snapshots of the flow field through the porous band at different Mach numbers, for $Re_{0}=450$. a),b): $Ma_{0}=0.0375$; c),d): $Ma_{0}=0.15$. The left panels (a,c) show the dimensionless velocity field $\left\lvert\bm{u}^{\star}\right\rvert=\left\lvert\bm{u}\right\rvert/U_{0}$. The right panels (b,d) show the dimensionless pressure field.\label{fig:porous_ts}}
\end{figure}

Finally, we increase the inflow Mach number to $Ma_{0}=0.15$, which results in transonic flow through the porous matrix; regions with local Mach number greater than unity occur within the pore space. The initial transients in the flow field take longer to decay (they last until $t^{\star}\approx300$), as the flow is choked - increases in the upstream pressure as the inflow boundary tracks the desired velocity are unable to force the flow through the porous matrix at a faster rate. Figure~\ref{fig:porous_ts} shows the dimensionless velocity (left) and pressure (right) fields with $Re_{0}=450$, for $Ma_{0}=0.0375$ (upper panels - a,b) and $Ma_{0}=0.15$ (lower panels - c,d). In the lower panels, regions with $\left\lvert\bm{u}^{\star}\right\rvert>6.\dot{6}$ correspond to local Mach numbers above unity. We see that for the transonic case, the velocity magnitude is greater through the porous band, with velocities approaching $10U_{0}$, corresponding to local Mach numbers of $1.5$. There is a larger pressure drop through the porous band for the transonic case, with increased unsteady flow away from the porous region. Within the porous region, local shock waves are visible.

Although this test problem is somewhat idealised, it provides a demonstration of the potential of LABFM. We are able to conduct high order ($4^{th}$ order at boundaries, $6^{th}$ internally) simulations of low and moderate Mach number flows at moderate Reynolds numbers (up to $Re_{p}=968$ here), in complex geometries. Discretisation of the domain is trivial. Although data is not available to rigorously validate the transonic flow field, the ability of LABFM to remain stable in the presence of steep gradients and incipient shocks provides a good demonstration of the robustness of the method, and in particular the filtering technique introduced in Section~\ref{filters}. In future, we intend to extend this work to enable three-dimensional simulations in geometries where the generation of high-quality body-conforming meshes is prohibitively resource intensive. This will allow us to investigate the behaviour of turbulence in a geometrically realistic porous matrix.

\section{Conclusions}\label{conc}

In this work we have presented several new developments for the Local Anistropic Basis Function method (LABFM), which allow numerical simulations of the isothermal Navier-Stokes equations in complex geometries. The main numerical developments are as follows:
\begin{enumerate}
\item We introduced anisotropic basis functions (ABFs) based on orthogonal polynomials. These provide a significant improvement in the condition number of the local linear systems which form the heart of LABFM, and therefore reduce the limiting residual error associated with the method, making extremely high order (e.g. tenth) practical. We also demonstrated the ability of LABFM yield high order gradient and Laplacian operators on node distributions with spatially varying resolution.
\item We introduced a stencil optimisation procedure, which improves the wavenumber response of the discrete operators, whilst reducing computational costs. The optimisation procedure here is relatively straightfoward, but still provides computational savings of $17.5\%$, whilst improving the resolving power of the discrete operators. We propose that a future avenue of research will be to develop more sophisticated optimisation procedures, in which the ABFs are adjusted alongside the symmetry of the stencil to further reduce the computational costs, and remove numerical dissipation from the operators.
\item To overcome the instabilities occuring in collocated schemes, we introduced a filtering technique, adapted from finite difference methods. In our approach, the damping coefficients are automatically calculated to ensure the filter satisfies some user-defined dealiasing rule. This approach appears to be more robust than the inclusion of hyperviscosity in the governing equations, with the further advantage that no additional time-step constraint is introduced. This filtering technique allows stable simulations of the Navier Stokes equations, with up to $9^{th}$ order for unbounded domains demonstrated herein. In this work, we studied the isothermal compressible Navier Stokes equations, but the approach also provides stability for fractional step schemes for incompressible flows. Two further studies are planned, extending the method to simulate a) incompressible, and b) fully compressible (i.e. with an energy equation) flows.
\item We introduced a characteristic-based boundary condition framework which allows for accurate simulations of flows with complex geometries. The framework involves the use of a combination of one-dimensional LABFM and $5$-point finite difference stencils, and yields $4^{th}$ order accurate boundary conditions for inflows, outflows and walls. Furthermore, it allows us to simulate solid boundaries with complex geometries, provided the boundary curvature is adequately resolved. The development of LABFM to yield stable, high order, one-sided differences is an active area of research for us, which if realised will provide further geometric flexibility.
\end{enumerate}

% Porous
With the above developments in place, we demonstrated the ability of our method to simulate subsonic flows through the complex geometry of an inhomogeneous porous matrix at moderate Reynolds numbers (up to $Re_{p}=968$). These results show the potential of LABFM to enable highly accurate simulations of flows in complex geometries, whilst preliminary results for transonic flows indicate the robustness of the method.

%% Eulerian/Lagrangian
Though the methods presented herein are Eulerian, there is no fundamental obstacle to their implementation in a Lagrangian framework, with node positions updated through the same time-integration scheme as the Navier-Stokes equations. The primary obstacle to a Lagrangian approach is the computational cost associated with the evaluation and solution of consistency matrices for each node at each time-step. The development of techniques to accelerate such implementations is an active area of research for our group.

%% Big picture
LABFM has accuracy and stability properties akin to high order finite differences, whilst retaining the geometric flexibility of many other mesh-free methods, and we have shown that these properties can be exploited to obtain a powerful numerical tool for simulations in complex geometries. Future developments, including an extension of this work to three-dimensional simulations, will enable direct numerical simulations of turbulence in realistic geometries with hitherto unparalleled fidelity.

\section*{Acknowledgements}
We are grateful for financial support from the Leverhulme Trust, via Research Project Grant RPG-2019-206. We are grateful to members of the University of Manchester SPH research group for many useful discussions. 

\appendix
\section{One-dimensional LABFM}\label{1dlabfm}

Near boundaries we use a one-dimensional form of LABFM to calculate derivatives in the direction tangential to the boundary. One-dimensional LABFM follows the same principles as LABFM in two dimensions. For completeness we present the details of the implementation here. For a given node $i$, the set of nodes which lie along the same boundary-tangential row of the locally orthogonal node distribution (see Figure~\ref{fig:stencils}), and also within the computational stencil of node $i$, are denoted $B_{i}$. The coordinate along the boundary tangent direction is $\eta$, and $\eta_{ji}=\bm{r}_{ji}\cdot\bm{t}_{i}$, where $\bm{t}_{i}$ is the unit vector tangential to the boundary.
We begin by defining a vector of monomials
\begin{equation}\bm{H}_{ji}=\begin{bmatrix}\eta_{ji},&\frac{\eta_{ji}^{2}}{2},\frac{\eta_{ji}^{3}}{6},&\frac{\eta_{ji}^{4}}{24},\dots\frac{\eta_{ji}^{m}}{m!}\end{bmatrix}^{T},\end{equation}
and a vector operator of partial derivatives as
\begin{equation}\bm{D}\left(\cdot\right)=\begin{bmatrix}\frac{\partial\left(\cdot\right)}{\partial{\eta}},&
\frac{\partial^{2}\left(\cdot\right)}{\partial{\eta}^{2}},&\frac{\partial^{3}\left(\cdot\right)}{\partial{\eta}^{3}},&\frac{\partial^{4}\left(\cdot\right)}{\partial{\eta}^{4}},\dots,\frac{\partial^{m}\left(\cdot\right)}{\partial{\eta}^{m}}\end{bmatrix}^{T},\end{equation}
where $m$ is the desired order of polynomial reproduction. Note the vector $\bm{H}_{ji}$ has length $n=m$. The discrete operator on node $i$ is defined as
\begin{equation}L^{d}_{i}\left(\cdot\right)=\displaystyle\sum_{j\in{B}_{i}}\left(\cdot\right)_{ji}w^{d}_{ji}\label{eq:general_do1d}\end{equation}
Equation~\eqref{eq:general_do1d} approximates $\left.\bm{C^{d}}\cdot\bm{D}\left(\cdot\right)\right\rvert_{i}$ where $\bm{C^{d}}$ is defined as
\begin{equation}\bm{C^{d}}=\begin{cases}\begin{bmatrix}1,&0,&0,&0,&0,&0,&0,&0\dots\end{bmatrix}^{T}&\text{if }d=\eta\\
\begin{bmatrix}0,&1,&0,&0,&0,&0,&0,&0\dots\end{bmatrix}^{T}&\text{if }d=\eta^{2}.\end{cases}\end{equation}
As for the two-dimensional case, we set $w^{d}_{ji}=\bm{W}_{ji}\cdot\bm{\Psi^{d}}_{i}$, where here the basis functions are functions of $\eta$ only: $W_{ji}=W\left(\eta_{ji}/h_{i}\right)$. We construct the linear system
\begin{equation}\bm{M}_{i}=\displaystyle\sum_{j\in{B}_{i}}\bm{H}_{ji}\otimes\bm{W}_{ji},\label{eq:disc_form1d}\end{equation}
and then obtain the weights $\bm{\Psi^{d}}_{i}$ by solving the system
\begin{equation}\bm{M}_{i}\bm{\Psi^{d}}_{i}=\bm{C^{d}}.\label{eq:lsys1d}\end{equation}
Having solved~\eqref{eq:lsys1d} we calculate  the weights $w^{d}_{ji}=\bm{W}_{ji}\cdot\bm{\Psi^{d}}_{i}$. In the present work we use ABFs based on univariate Hermite polynomials, where the $q$-th ABF is defined as
\begin{equation}W_{ji}^{q}=\frac{\psi\left(\lvert{\eta}_{ji}\rvert/h_{i}\right)}{\sqrt{2^{q}}}H_{q}\left(\frac{\eta_{ji}}{h_{i}\sqrt{2}}\right),\end{equation}
in which $H_{q}$ is the $q$-th order univariate physicists Hermite polynomial, and $\psi$ is an RBF.

\bibliographystyle{elsarticle-num-names}
\bibliography{jrckbib}

\end{document}